\documentclass[
superscriptaddress,
nofootinbib, twocolumn,
amsmath,amssymb,aps,
floatfix, prc]{revtex4-2}
\usepackage{graphicx} % Required for inserting images
\usepackage{amsmath}
\usepackage{amssymb}
\usepackage{bm}
\usepackage[dvipsnames]{xcolor}
\usepackage[colorlinks=true, pdfstartview=FitV, linkcolor=RedOrange, citecolor=Emerald, urlcolor=Cerulean]{hyperref}
\usepackage[utf8]{inputenc}
\usepackage{wrapfig}
\usepackage{tikz}
\usepackage{comment}
\usepackage[colorinlistoftodos]{todonotes}

\newcommand{\p}{\bm{p}}
\newcommand{\snn}{\sqrt{s_\mathrm{NN}}}
\newcommand{\xscape}{\textsc{x-scape}}
\newcommand{\xscapegim}{\textsc{x-scape-gim}}
\newcommand{\jetscape}{\textsc{jetscape}}
\newcommand{\Glb}{\textsc{3d-Glauber}}
\newcommand{\MUSIC}{\textsc{music}}

\newcommand{\pp}{$p$-$p$}
\renewcommand{\AA}{$A$-$A$}
\newcommand{\pA}{$p$-$A$}
\newcommand{\pPb}{$p$-$Pb$}
\newcommand{\matter}{\textsc{matter}}
\newcommand{\imatter}{\textsc{I-matter}}

\usetikzlibrary{shapes.geometric,arrows.meta,decorations.text,decorations.pathreplacing,calligraphy}

\tikzstyle{startstop} = [rectangle, rounded corners, 
minimum width=3cm, 
minimum height=1cm,
text width=3cm,
text centered, 
draw=black, fill=blue!30]
\tikzstyle{pythia} = [rectangle, rounded corners, 
minimum width=3cm, 
minimum height=1cm,
text centered, 
draw=black, 
fill=red!30]
\tikzstyle{soft} = [rectangle, rounded corners,
minimum width=3cm, 
minimum height=1cm, text centered, 
text width=3cm,
draw=black, fill=blue!30]

\tikzstyle{process} = [rectangle, rounded corners,
minimum width=3cm, 
minimum height=1cm, 
text centered, 
text width=3cm, 
draw=black, 
fill=orange!30]

\tikzstyle{had} = [rectangle, rounded corners, 
minimum width=3cm, 
minimum height=1cm, 
text centered, 
text width=3cm, 
draw=black, 
fill=blue!30!green!30]
\tikzstyle{decision} = [diamond, 
minimum width=3cm, 
minimum height=1cm, 
text centered, 
text width=2cm,
draw=black, 
fill=green!30]
\tikzstyle{arrow} = [thick,->,>=stealth]

\begin{document}
\title{A soft-hard framework with exact four momentum conservation for small systems }

%%%%%% AUTHORS %%%%%%%%%%%%%
%%% JETSCAPE Author List

\author{I.~Soudi}
\affiliation{Department of Physics and Astronomy, Wayne State University, Detroit MI 48201.}
\affiliation{University of Jyväskylä, Department of Physics, P.O. Box 35, FI-40014 University of Jyväskylä, Finland.}
\affiliation{Helsinki Institute of Physics, P.O. Box 64, FI-00014 University of Helsinki, Finland.}

\author{W.~Zhao}
\affiliation{Department of Physics and Astronomy, Wayne State University, Detroit MI 48201.}
\affiliation{Department of Physics, University of California, Berkeley CA 94270.}
\affiliation{Nuclear Science Division, Lawrence Berkeley National Laboratory, Berkeley CA 94270.}

\author{A.~Majumder}
\affiliation{Department of Physics and Astronomy, Wayne State University, Detroit MI 48201.}

\author{C.~Shen}
\affiliation{Department of Physics and Astronomy, Wayne State University, Detroit MI 48201.}
\affiliation{RIKEN BNL Research Center, Brookhaven National Laboratory, Upton NY 11973.}

\author{J.~H.~Putschke}
\affiliation{Department of Physics and Astronomy, Wayne State University, Detroit MI 48201.}

\author{B.~Boudreaux}
\affiliation{Department of Physics and Astronomy, Wayne State University, Detroit MI 48201.}

%%%%%%%%%%%%%%%%%%%%%%%%%%%%%%%%
\author{A.~Angerami}
\affiliation{Lawrence Livermore National Laboratory, Livermore CA 94550.}

\author{R.~Arora}
\affiliation{Department of Computer Science, Wayne State University, Detroit MI 48202.}

\author{S.~A.~Bass}
\affiliation{Department of Physics, Duke University, Durham, NC 27708, USA}

\author{Y.~Chen}
\affiliation{Laboratory for Nuclear Science, Massachusetts Institute of Technology, Cambridge MA 02139.}
\affiliation{Department of Physics, Massachusetts Institute of Technology, Cambridge MA 02139.}
\affiliation{Department of Physics and Astronomy, Vanderbilt University, Nashville TN 37235.}

\author{R.~Datta}
\affiliation{Department of Physics and Astronomy, Wayne State University, Detroit MI 48201.}

\author{L.~Du}
\affiliation{Department of Physics, McGill University, Montr\'{e}al QC H3A\,2T8, Canada.}
\affiliation{Department of Physics, University of California, Berkeley CA 94270.}
\affiliation{Nuclear Science Division, Lawrence Berkeley National Laboratory, Berkeley CA 94270.}

\author{R.~Ehlers}
\affiliation{Department of Physics, University of California, Berkeley CA 94270.}
\affiliation{Nuclear Science Division, Lawrence Berkeley National Laboratory, Berkeley CA 94270.}

\author{H.~Elfner}
\affiliation{GSI Helmholtzzentrum f\"{u}r Schwerionenforschung, 64291 Darmstadt, Germany.}
\affiliation{Institute for Theoretical Physics, Goethe University, 60438 Frankfurt am Main, Germany.}
\affiliation{Frankfurt Institute for Advanced Studies, 60438 Frankfurt am Main, Germany.}

\author{R.~J.~Fries}
\affiliation{Cyclotron Institute, Texas A\&M University, College Station TX 77843.}
\affiliation{Department of Physics and Astronomy, Texas A\&M University, College Station TX 77843.}

\author{C.~Gale}
\affiliation{Department of Physics, McGill University, Montr\'{e}al QC H3A\,2T8, Canada.}

\author{Y.~He}
%\affiliation{Guangdong Provincial Key Laboratory of Nuclear Science, Institute of Quantum Matter, South China Normal University, Guangzhou 510006, China.}
%\affiliation{Guangdong-Hong Kong Joint Laboratory of Quantum Matter, Southern Nuclear Science Computing Center, South China Normal University, Guangzhou 510006, China.}
\affiliation{School of Physics and Optoelectronics, South China University of Technology, Guangzhou 510640, China}

\author{B.~V.~Jacak}
\affiliation{Department of Physics, University of California, Berkeley CA 94270.}
\affiliation{Nuclear Science Division, Lawrence Berkeley National Laboratory, Berkeley CA 94270.}

\author{P.~M.~Jacobs}
\affiliation{Department of Physics, University of California, Berkeley CA 94270.}
\affiliation{Nuclear Science Division, Lawrence Berkeley National Laboratory, Berkeley CA 94270.}

\author{S.~Jeon}
\affiliation{Department of Physics, McGill University, Montr\'{e}al QC H3A\,2T8, Canada.}

\author{Y.~Ji}
\affiliation{Department of Statistical Science, Duke University, Durham NC 27708.}

\author{L.~Kasper}
\affiliation{Department of Physics and Astronomy, Vanderbilt University, Nashville TN 37235.}

\author{M.~Kelsey}
\affiliation{Department of Physics and Astronomy, Wayne State University, Detroit MI 48201.}

\author{M.~Kordell~II}
\affiliation{Cyclotron Institute, Texas A\&M University, College Station TX 77843.}
\affiliation{Department of Physics and Astronomy, Texas A\&M University, College Station TX 77843.}

\author{A.~Kumar}
\affiliation{Department of Physics, University of Regina, Regina, SK S4S 0A2, Canada.}
\affiliation{Department of Physics, McGill University, Montr\'{e}al QC H3A\,2T8, Canada.}

\author{R.~Kunnawalkam-Elayavalli}
\affiliation{Department of Physics and Astronomy, Vanderbilt University, Nashville TN 37235.}

\author{J.~Latessa}
\affiliation{Department of Computer Science, Wayne State University, Detroit MI 48202.}

\author{Y.-J.~Lee}
\affiliation{Laboratory for Nuclear Science, Massachusetts Institute of Technology, Cambridge MA 02139.}
\affiliation{Department of Physics, Massachusetts Institute of Technology, Cambridge MA 02139.}

\author{R.~Lemmon}
\affiliation{Daresbury Laboratory, Daresbury, Warrington, Cheshire, WA44AD, United Kingdom.}

\author{M.~Luzum}
\affiliation{Instituto  de  F\`{i}sica,  Universidade  de  S\~{a}o  Paulo,  C.P.  66318,  05315-970  S\~{a}o  Paulo,  SP,  Brazil. }

\author{S.~Mak}
\affiliation{Department of Statistical Science, Duke University, Durham NC 27708.}

\author{A.~Mankolli}
\affiliation{Department of Physics and Astronomy, Vanderbilt University, Nashville TN 37235.}

\author{C.~Martin}
\affiliation{Department of Physics and Astronomy, University of Tennessee, Knoxville TN 37996.}

\author{H.~Mehryar}
\affiliation{Department of Computer Science, Wayne State University, Detroit MI 48202.}

\author{T.~Mengel}
\affiliation{Department of Physics and Astronomy, University of Tennessee, Knoxville TN 37996.}

\author{C.~Nattrass}
\affiliation{Department of Physics and Astronomy, University of Tennessee, Knoxville TN 37996.}

\author{J.~Norman}
\affiliation{Oliver Lodge Laboratory, University of Liverpool, Liverpool, United Kingdom.}

\author{C.~Parker}
\affiliation{Cyclotron Institute, Texas A\&M University, College Station TX 77843.}
\affiliation{Department of Physics and Astronomy, Texas A\&M University, College Station TX 77843.}

\author{J.-F.~Paquet}
\affiliation{Department of Physics and Astronomy, Vanderbilt University, Nashville TN 37235.}

\author{H.~Roch}
\affiliation{Department of Physics and Astronomy, Wayne State University, Detroit MI 48201.}

\author{G.~Roland}
\affiliation{Laboratory for Nuclear Science, Massachusetts Institute of Technology, Cambridge MA 02139.}
\affiliation{Department of Physics, Massachusetts Institute of Technology, Cambridge MA 02139.}

\author{B.~Schenke}
\affiliation{Physics Department, Brookhaven National Laboratory, Upton NY 11973.}

\author{L.~Schwiebert}
\affiliation{Department of Computer Science, Wayne State University, Detroit MI 48202.}

\author{A.~Sengupta}
\affiliation{Cyclotron Institute, Texas A\&M University, College Station TX 77843.}
\affiliation{Department of Physics and Astronomy, Texas A\&M University, College Station TX 77843.}

\author{M.~Singh}
\affiliation{Department of Physics and Astronomy, Vanderbilt University, Nashville TN 37235.}

\author{C.~Sirimanna}
\affiliation{Department of Physics and Astronomy, Wayne State University, Detroit MI 48201.}
\affiliation{Department of Physics, Duke University, Durham, NC 27708, USA}

\author{D.~Soeder}
\affiliation{Department of Physics, Duke University, Durham NC 27708.}

\author{R.~A.~Soltz}
\affiliation{Department of Physics and Astronomy, Wayne State University, Detroit MI 48201.}
\affiliation{Lawrence Livermore National Laboratory, Livermore CA 94550.}

\author{Y.~Tachibana}
\affiliation{Akita International University, Yuwa, Akita-city 010-1292, Japan.}

\author{J.~Velkovska}
\affiliation{Department of Physics and Astronomy, Vanderbilt University, Nashville TN 37235.}

\author{G.~Vujanovic}
\affiliation{Department of Physics, University of Regina, Regina, SK S4S 0A2, Canada.}

\author{X.-N.~Wang}
\affiliation{Key Laboratory of Quark and Lepton Physics (MOE) and Institute of Particle Physics, Central China Normal University, Wuhan 430079, China.}
\affiliation{Department of Physics, University of California, Berkeley CA 94270.}
\affiliation{Nuclear Science Division, Lawrence Berkeley National Laboratory, Berkeley CA 94270.}

\author{X.~Wu}
\affiliation{Department of Physics, McGill University, Montr\'{e}al QC H3A\,2T8, Canada.}
\affiliation{Department of Physics and Astronomy, Wayne State University, Detroit MI 48201.}

\collaboration{The JETSCAPE Collaboration}

\begin{abstract}
   A new framework, called \xscape, for the combined study of both hard and soft transverse momentum sectors in high energy proton-proton (\pp) and proton-nucleus (\pA) collisions is set up. A dynamical initial state is set up using the \Glb{} model with transverse locations of hotspots within each incoming nucleon. A hard scattering that emanates from two colliding hotspots is carried out using the Pythia generator. Initial state radiation from the incoming hard partons is carried out in a new module called \imatter, which includes the longitudinal location of initial splits. The energy-momentum of both the initial hard partons and their associated beam remnants is removed from the hot spots, depleting the energy-momentum available for the formation of the bulk medium. Outgoing showers are simulated using the \matter{} generator, and results are presented for both cases, allowing for and not allowing for energy loss. First comparisons between this hard-soft model and single inclusive hadron and jet data from \pp{} and minimum bias \pPb{} collisions are presented. Single hadron spectra in \pp{} 
are used to carry out a limited (in number of parameters) Bayesian calibration of the model. Fair comparisons with data are indicative of the utility of this new framework. Theoretical studies of the correlation between jet $p_T$ and event activity at mid and forward rapidity are carried out.
\end{abstract}
\date{\today}

\maketitle

\section{Introduction}

At the start of the heavy-ion program at the Relativistic Heavy-Ion Collider (RHIC), highly asymmetric collisions, such as proton-nucleus \pA, or deuteron-nucleus $d$-$A$, were considered as a baseline to the central topic of nucleus-nucleus \AA~collisions, especially for events with hard interactions~\cite{STAR:2003pjh,PHENIX:2003qdw}. 
The energy deposition (or stopping) in these collisions was not considered sufficient for the production of a quark-gluon plasma (QGP). Indeed, no significant final state modification of high transverse momentum (high-$p_T$) particles was measured in these collisions, compared to the binary collision enhanced yield in \pp{} collisions.\footnote{{In this paper, we denote a high-$p_T$ particle (or jet) as one which possesses a $p_T\gtrsim 8$~GeV ($\gtrsim 20$~GeV).}} The lack of final state modification allowed for a study of the initial state effects such as nuclear shadowing~\cite{Arnold:1983mw} and the Cronin effect~\cite{Cronin:1973fd,Cronin:1974zm}. These measurements, in turn, allowed for a more accurate quantification of final state effects.

This picture changed in 2012, when final state collective behavior was identified in high multiplicity \pA~\cite{ATLAS:2012cix,CMS:2012qk,ALICE:2012eyl} and even in \pp~\cite{CMS:2012qk} collisions (at $\sqrt{s_{\rm NN}} \gtrsim 5$~TeV) at the Large Hadron Collider (LHC). It would seem that it is indeed possible to produce a QGP, even in the small systems created in \pp{} and \pPb{} collisions~\cite{Nagle:2018nvi, Shen:2020gef, Noronha:2024dtq, Grosse-Oetringhaus:2024bwr}.
Hydrodynamic simulations extrapolated from calibrated models in large heavy-ion collisions can quantitatively describe the anisotropic flow measurements in small systems \cite{Weller:2017tsr, Schenke:2020mbo}. These calculations demonstrated that particle momentum anisotropy in high multiplicity collisions is dominated by final-state interactions~\cite{Schenke:2019pmk}. Furthermore, thermal electromagnetic radiation can serve as an independent signature of the QGP formation in small systems~\cite{Shen:2015qba, Shen:2016zpp, Khachatryan:2020rqr}.

{The physical picture underlying small systems became further involved with the observation of a large suppression in the nuclear modification factor ($R_{pA}$) for jets in ``central" events with $p_T$ significantly exceeding $100$~GeV, in concert with an \emph{equal} enhancement for similar jets in ``peripheral" events at the ATLAS detector~\cite{ATLAS:2014cpa}. The suppression and enhancement were balanced such that no modification was discerned in the minimum bias spectrum at high $p_T$. 
In Ref.~\cite{ATLAS:2014cpa}, centrality was deduced by relating measurements of transverse energy (or event activity) produced at forward rapidities between $-4.9 \leq \eta \leq -3.2$, in the $Pb$ going direction, with the geometry of the collision, using a wounded nucleon model~\cite{Bialas:1976ed}. While the relation between forward transverse energy and the mean number of binary collisions, in a given centrality bin, is not straightforward in $p$-$A$ collisions, these measurements do clearly demonstrate a clear (and surprising) negative correlation between the energy of hard scattering and the transverse energy or event activity at forward rapidity, for jet $p_T$'s that  approach and significantly exceed $100$~GeV. }

{An extensive study by the ALICE collaboration~\cite{ALICE:2014xsp} found various correlations between the $p_T$ of the hard process, and the event activity measured in detectors placed at forward rapidity ($2.8 < \eta < 5.1$ on the $Pb$ going side, and $1.7 < \eta < 3.7$ on the $p$ going side, as well as in zero degree calorimeters placed close to beam rapidity). 
Depending on the detector used and the relation between measurements and the deduced number of binary nucleon-nucleon collisions, one observes a positive, negative, or no correlation between the nuclear modification factor of charged hadrons and the putative centrality of the collision.  
However, ALICE hadron measurements in Ref.~\cite{ALICE:2014xsp} do not exceed a $p_T$ of $30$~GeV. }

{The positive correlation between hadrons with $8$~GeV~$\lesssim p_T \leq 20$~GeV and the event activity at forward rapidity was partially corroborated by the ATLAS collaboration in Ref.~\cite{ATLAS:2016xpn}. However, the relation between event activity and the calculated number of binary collisions, as required for an $R_{pA}$ measurement, was found to be strongly dependent on the theoretical model used. 
Ref.~\cite{ATLAS:2016xpn} also suffered from the lack of a measured \pp{} reference at $\sqrt{s_{\rm NN}} = 5.02$~TeV.}

Later measurements of charged hadrons by the ALICE collaboration, with $6$~GeV $\leq p_T \leq 50$~GeV, in Ref.~\cite{ALICE:2017svf}, show a positive correlation with event activity, measured at forward rapidities ($2.8 \leq \eta \leq 5.1$ in $Pb$ going side). However, no clear correlation of event activity with the $p_T$ of jets,  triggered against hadrons with $6$~GeV $\leq p_T \leq 7$~GeV (or $12$~GeV$\leq p_T \leq 50$~GeV), is observed. The reader should note that measured jet energies in ALICE are limited to 
$p_T \leq 120$~GeV, and thus do not probe the very high energy jet range explored by the ATLAS measurements in Ref.~\cite{ATLAS:2014cpa}.

{Recent measurements by the ATLAS detector~\cite{ATLAS:2022iyq} in this intermediate $p_T$ range, using single inclusive and jet triggered hadrons with $p_T \leq 60$~GeV (at midrapidity), showed no correlation with event activity, measured near beam rapidity, using zero degree calorimeters. In this measurement, jet energies were limited to $p_T^{jet} \gtrapprox 30$~GeV or $p_T^{jet} \gtrapprox 60$~GeV (consistent with energies measured in ALICE and much lower than the jet energies explored by the previous ATLAS measurements in Ref.~\cite{ATLAS:2014cpa}).}

In spite of evidence for the formation of a QGP in high multiplicity (or high event activity) events in \pA{} collisions, none of the above measurements by the ALICE or ATLAS collaborations, containing a high-$p_T$ particle or jet, find evidence for a non-negligible amount of energy loss, with bounds placed at $\Delta E \lesssim 1\%$~\cite{ATLAS:2022iyq}, or $p_T \lesssim 0.4$~GeV out of jet cone~\cite{ALICE:2017svf}.
The lack of significant energy loss in \pA{} collisions is further complicated by the observation of an elliptic anisotropy at high $p_T$ in both \pp~\cite{ATLAS:2016yzd} and \pPb{}~\cite{ATLAS:2019vcm} collisions.

The wide range of experimental studies mentioned above, paint a complex picture of the dynamics of \pA{} collisions, specifically collisions with larger event activity. Is a QGP formed in high multiplicity (event activity) \pA{} collisions? Can events with high multiplicity (event activity) be produced entirely due to bulk dynamics, or are these generated by hard events? Is there a measurable signal of energy loss in these systems? To systematically answer all these questions, one requires an event generator framework that can simultaneously simulate hard scattering with both initial and final state radiation, and bulk evolution, including a possible QGP phase. The simulator should reliably simulate a wide range in rapidity to allow for detailed studies of the event activity both with and without hard processes~\cite{Pierog:2013ria, Kanakubo:2021qcw}.
Such efforts should be differentiated from those that do not include a macroscopic fluid dynamical simulation, e.g. \cite{Bierlich:2018xfw}.
A modular and extensible event generator framework would provide insight into the observed elliptic anisotropy for high $p_T$ hadrons, in tandem with the lack of any significant energy loss for high $p_T$ hadrons and jets.

In central collisions of $Pb$ on $Pb$, a large fraction of the 208 nucleons in $Pb$ participate in the deposition of energy, leading to the formation of an extended QGP~\cite{ALICE:2010khr}. In almost head-on collision systems, the produced QGP can have a transverse radius comparable to the radius of a $Pb$ nucleus~\cite{Shen:2011eg}. Hard jets lose energy by scattering and radiation in this dense, extended environment~\cite{Majumder:2010qh,Cao:2020wlm,JETSCAPE:2022jer}. Jets are formed in the hard scattering of two incoming partons from two specific nucleons. While the formation of hard partons within these nucleons may result in either an excess or depletion of the energy available for QGP formation, this has negligible effect on the bulk dynamics of the system, that receives contributions from hundreds of other, similar nucleon-nucleon interactions. 

In a small system, such as those created in a \pp{} or \pPb{} collisions, the production of a hard parton {can} affect the amount of energy deposited in the bulk sector, either increasing or decreasing it~\cite{Kordell:2016njg}. 
Thus, the bulk properties of small systems {may be} affected by the presence of a hard scattering. 
Also, the small size of the produced QGP in \pp, or even \pPb{}, barely larger than the transverse size of a proton, {is expected} to lead to minimal final state modification of the produced jets and leading hadrons~\cite{Huss:2020whe}. 
Thus, small systems provide unique laboratories to study the correlation between the hard (perturbative) and soft (non-perturbative) sectors of the initial state. 

Since the observation of collective behavior in small systems, several different effects have been ascribed to the high-$p_T$ measurements in Ref.~\cite{ATLAS:2014cpa}, including color transparency~\cite{Alvioli:2014eda}, initial state energy loss~\cite{Ovanesyan:2015dop}, to corrections to the final state energy loss formula from small path lengths~\cite{Kolbe:2015rvk}. To disentangle all these effects will require a systematic study with an extensible framework. 
One of the goals of the X-ion (collisions) with a Statistically and Computationally Advanced Program Envelope (\xscape) project is to provide such a framework~\cite{x-scape-github}. In this first paper using the \xscape{} framework, we explore the results from the minimal configuration, which only includes energy momentum conservation between the hard and soft sectors~\cite{Kordell:2016njg}, in \pp{} and \pPb{}, both with and without energy loss in the final state of \pPb{}. Any further effects such as color transparency, corrections to initial or final state energy loss, should be considered in addition to these basic constraints of energy momentum conservation. 

The \xscape{} project is the second undertaking of the JETSCAPE collaboration, extending the JETSCAPE framework~\cite{Putschke:2019yrg} to small systems and lower energy heavy-ion collisions. On completion, the \xscape{} framework will incorporate modular simulators of \pp, \AA, \pA~from beam energy scan energies at RHIC to top LHC energies and some aspects of $e$-$p$ and $e$-$A$ collisions relevant to EIC energies, within a single framework. The current work reports on the first study of high and low multiplicity \pp~and \pA~collisions. In particular, we will explore the consequences of exact energy-momentum conservation between the hard and soft sectors in these small systems. 

The remainder of the paper is organized as follows: In Sec.~\ref{sec:theory-framework}, we describe the overarching framework used to simulate \pp~and \pA~collisions, with exact energy-momentum conservation between the hard and soft sectors, as simulated by the new initial state modules of \Glb{} and the Initial-state-Modular-All-Twist-Transverse-and-Elastic-scattering-induced-Radiation (\imatter{}) module. 
In Sec.~\ref{sec:results}, we present the first results from this new framework where the final state shower in both \pp~and \pA~are carried out assuming no energy loss, i.e., vacuum shower. In Sec.~\ref{sec:energy-loss}, we study the effect of final state energy loss by carefully studying the evolution of the produced medium with respect to the propagating and developing jets, using the maximal rate of energy loss possible, using the \matter{} module without any coherence effects~\cite{MehtarTani:2011tz,CasalderreySolana:2012ef,Kumar:2019uvu,JETSCAPE:2022jer}. A summary and outlook to future work will be presented in Sec.~\ref{sec:summary}. 
{While interesting and somewhat related effects have been noticed in $d$-$A$ and \pA{} collisions at RHIC energies~\cite{STAR:2003pjh,PHENIX:2003qdw,PHENIX:2023dxl}, in this first effort using the new \xscape{} framework, we will only present comparisons with LHC data. }

\section{The Theoretical Framework}
\label{sec:theory-framework}

The underlying picture which animates the new modules simulating \pp~and \pA, within the \xscape{} framework, is the existence of short distance sub-nucleon fluctuations within the proton~\cite{Gale:2012rq}. The majority of these sub-nucleon fluctuations, or hot spots, have inverse distance scales, which are $\sim 1$~GeV. In most collisions, these will engender soft interactions with hot spots from the opposing nucleon(s). Energy redistribution from these interactions, leading to bulk dynamics and soft (low $p_T$) particle production, will be treated using non-perturbative models. Occasionally, fluctuations will yield a hard parton, which may undergo initial state gluon bremsstrahlung before hard scattering off another hard parton from the opposing nucleon(s), leading to the production of partons with a large $p_T$ and final state radiation. All of these processes are treated using models based on pQCD. 
However, producing a hard parton will necessarily deplete the energy of the nearby hot spot in the proton. This hot spot will now no longer have its full energy available to contribute to bulk particle production via soft exchanges with hot spots in the opposing nucleon(s). Also, any transverse momentum of  the hard parton should be balanced by an opposing transverse momentum in the hot spot. In actual simulations, the beam remnant, typically a di-quark, also draws energy and momentum from the same hot spot.

%%%%%%%%%%%%%%%%%%%%%%%%%%%%%%%
% Flowchart 
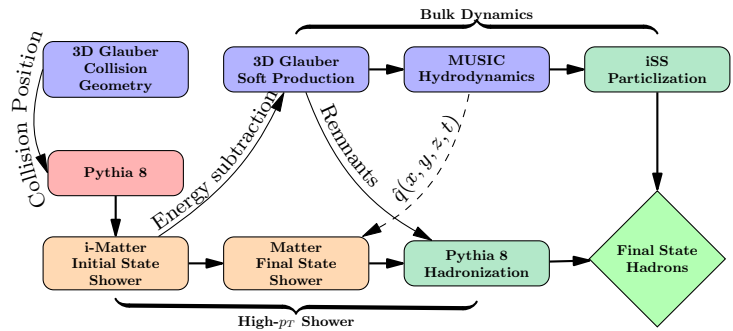
\begin{figure*}%[b!]
    \centering
    \begin{tikzpicture}[node distance=3cm,>={Stealth[inset=0pt,length=8pt,angle'=28,round]}, scale=0.6, every node/.style={scale=0.6}]]

    %\node (PYT) [startstop] {Pythia 8};
    \node (3D1) [startstop] {\textbf{3D Glauber\\ Collision Geometry}};
    \node (PYT) [pythia, below of=3D1,yshift=0.7cm] {\textbf{Pythia 8}};
    \node (iMA) [process, below of=PYT, yshift=1cm] {\textbf{i-Matter\\ Initial State Shower}};
    \node (MAT) [process, right of=iMA,xshift=1cm] {\textbf{Matter\\ Final State Shower}};

    \node (3DG) [soft, right of=3D1, xshift=1cm] {\textbf{3D Glauber\\ Soft Production}};
    \node (MUS) [soft, right of=3DG, xshift=1cm] {\textbf{MUSIC\\ Hydrodynamics}}; 
     
    \node (iSS) [had, right of=MUS, xshift=1cm] {\textbf{iSS\\ Particlization}};
    \node (HAD) [had, right of=MAT, xshift=1cm] {\textbf{Pythia 8\\ Hadronization}};
    \node (FSH) [decision, right of=HAD, yshift= 0.1cm,xshift=1cm] {\textbf{Final State\\ Hadrons}};

    \draw [->, thick] (PYT) -- (iMA);
    \draw [->, thick] (iMA) -- (MAT);
    
    \draw [<-, dashed] (MAT) to [bend right=25] (MUS);
    \draw [decoration={raise=1.2ex, text along path, text format delimiters={<}{>}, text={<\footnotesize> {~~$\hat{q}(x,y,z,t)$}}, text align={center}},decorate] (MAT) to [bend right=25] (MUS) ;
    % \draw [->] (iMA) to [bend right=10] node[pos=0.5, sloped, anchor=north] {\small Energy subtraction} (3DG);
    \draw [->] (iMA) to [bend right=15]  (3DG);
    \draw [decoration={raise=0.5ex, text along path, text format delimiters={<}{>}, text={<\footnotesize>Energy subtraction},text align={center}},decorate]  (iMA) to [bend right=15]  (3DG);
    % \draw [->] (3D1.west) to [bend right=20] node[pos=0.5, anchor=south, rotate=180, sloped] {\small Collision Position} (PYT.west);
    \draw [<-] (PYT.west) to [bend left=20] (3D1.west) ;
    \draw [decoration={raise=0.7ex, text along path, text format delimiters={<}{>}, text={<\footnotesize>Collision Position}, text align={center}},decorate] ([yshift=-0.5cm, xshift=0.3cm]PYT.south west) to [bend left=25] ([yshift=0.5cm, xshift=0.3cm]3D1.north west) ;
    % \draw [->] (3DG) to [bend right=10] node[pos=0.5, sloped, anchor=south] {\small Remnants} (HAD);
    \draw [->] (3DG) to [bend right=20] (HAD);
    \draw [decoration={raise=0.7ex, text along path, text format delimiters={<}{>}, text={<\footnotesize> Remnants~~~~~~~~~}, text align={center}},decorate] (3DG) to [bend right=20] (HAD) ;
    \draw [->, thick] (3DG) -- (MUS);
    \draw [->, thick] (MAT) -- (HAD);
    \draw [->, thick] (MUS) -- (iSS);
    \draw [->, thick] (iSS) -- (FSH);
    \draw [->, thick] (HAD) -- (FSH);
    % Braces 
    \node (BLK) [above of=MUS,yshift=-1.8cm]{\textbf{Bulk Dynamics}};
    \draw [decorate, decoration = {calligraphic brace,raise=5pt,amplitude=5pt}, ultra thick] (3DG.north) --  (iSS.north);
    \node (HRD) [below of=MAT,yshift=1.7cm]{\textbf{High-$p_T$ Shower}};
    \draw [decorate, decoration = {calligraphic brace,raise=5pt,amplitude=5pt,mirror}, ultra thick] ([yshift=0.1cm]iMA.south) --  ([yshift=-0.05cm]HAD.south);
    \end{tikzpicture}
    \caption{The workflow of the \xscapegim{} event generator for small collision systems.}
    \label{Fig:framework}
\end{figure*}

The fragmentation of these hard processes leads to the production of hadrons with a range of momenta, including hard and soft hadrons. Thus, the final ensemble of hadrons from an event with hard and soft exchanges will include hadrons that originated in purely soft and also in hard processes. The need for energy-momentum conservation becomes clear at this point; adding a hard process on top of soft exchanges without accounting for the balance in energy-momentum, as typically done for the case of heavy-ion collisions, will lead to a continuous increase of the multiplicity of the collision with the transverse momentum of the hardest exchange. 

\xscape{} is a versatile Monte-Carlo event generator framework for \pp, \pA, and \AA~collisions. In this work, we focus on the detailed physics implementations that are crucial for small collision systems. This particular simulator will be referred to as \xscapegim{} (an acronym for \Glb{} + pythia + \imatter{} + \MUSIC{} + \matter{}). The overarching simulation workflow is summarized in Fig.~\ref{Fig:framework}. This workflow presents a multi-stage hybrid approach for soft and hard particle production in relativistic nuclear collisions. For small collision systems, one key ingredient is to impose event-by-event energy-momentum conservation at the initial state of the collisions.

In the \xscapegim{} simulator, particle showers produced in hard processes are simulated using perturbative QCD techniques. By propagating the hard scattering partons backward in time, we model the space-time structure of the initial-state radiation (ISR) and compute the initial energies and momenta of the partons that originated from the hot spots in the colliding nucleons. Then, the hard partons' energies and momenta are subtracted from the incoming collision energy and momentum event by event. The remaining energy and momentum in the system are attributed to soft particle production.

Particle production with transverse momentum below 2 GeV is predominantly from non-perturbative processes in relativistic nuclear collisions. In the \xscapegim{} simulator, hot spots not affected by hard processes may interact with hot spots from opposing nucleon(s), producing strings that deposit energy, momentum, and baryon number. This provides the initial state of the energy-momentum tensor, which is further evolved according to the relativistic viscous hydrodynamic equation of motion. As the system evolves to a dilute region, individual fluid cells are converted into hadrons and propagated with hadronic scatterings and decays {(while this aspect is easily simulated with the SMASH generator~\cite{SMASH:2016zqf}, in this first paper using the \xscape{} framework, we will not invoke hadronic rescattering, in the interest of expediency)}. Compared with other high-energy event generators, the \xscapegim{} simulator provides a detailed space-time structure of the collision event. It builds in non-trivial correlations between soft and hard particle production with event-by-event energy-momentum conservation. In the following subsections, we will explain individual physics models following the workflow in Fig.~\ref{Fig:framework}.

\subsection{Collision Geometry}

The \xscapegim{} simulator employs the \Glb{} generator as the initial-state model for simulating the (3+1)D dynamics at the nuclear impact of the colliding proton/nuclei. 
Before the collision, the \Glb{} model samples the nucleon positions inside the nucleus according to the Woods-Saxon distributions or using the pre-generated configurations with realistic nuclear potentials. Individual nucleons have a Gaussian profile in the transverse plane,
\begin{equation}
    T_n(\textbf{r}_\perp) = \frac{1}{2 \pi B_G} e^{-r_\perp^2/(2B_G)},
    \label{eq:nucleonProfile}
\end{equation}
where the parameter $B_G$ controls the transverse size of the nucleon. 
Individual nucleon-nucleon collisions are determined by their impact parameter $\textbf{b}_\perp$, with the probability~\cite{Heinz:2011mh, Shen:2014vra},
\begin{equation}
    P(\textbf{b}_\perp) = 1 - \exp(-\sigma_{gg} T_{nn}(\textbf{b}_\perp)),
\end{equation}
where $\sigma_{gg}$ is the effective gluon-gluon cross section and $T_{nn}$ is the nucleon-nucleon overlapping function,
\begin{equation}
    T_{nn}(\textbf{b}_\perp) = \int d^2 \textbf{r}_\perp T_n(\textbf{r}_\perp) T_n(\textbf{r}_\perp - \textbf{b}_\perp) = \frac{e^{-b_\perp^2/(4B_G)}}{4 \pi B_G}.
\end{equation}
We compute the effective gluon-gluon cross section $\sigma_{gg}$ to ensure the inelastic nucleon-nucleon cross section $\sigma^\mathrm{inel}_\mathrm{NN}(\snn)$ is reproduced~\cite{Heinz:2011mh},
\begin{equation}
    \int d^2 \textbf{b}_\perp P(\textbf{b}_\perp) = \sigma^\mathrm{inel}_\mathrm{NN}(\snn).
\end{equation}

Unlike the Trento model~\cite{Moreland:2014oya}, our implementation ensures the inelastic cross sections in \pp{}, \pA{}, and \AA{} collisions are independent of the nucleon transverse size $B_G$. Table~\ref{tab:inelCrossX} shows the inelastic cross sections for a few collision systems at the top RHIC and LHC energies.
\begin{table}[h!]
    \centering
    \caption{The inelastic cross sections for \pp{}, \pA{}, and \AA{} collisions at the top RHIC and LHC energies from the \Glb{} model.}
    \begin{tabular}{c|c}
    \hline \hline
       collision system  &  $\sigma^\mathrm{inel}$ (b) \\ \hline
       p+p @ 200 GeV  &  0.042 \\ \hline
       p+Au @ 200 GeV & 1.75 \\ \hline
       Au+Au @ 200 GeV & 6.89 \\ \hline
       p+p @ 5020 GeV & 0.067 \\ \hline
       p+Pb @ 5020 GeV & 2.16 \\ \hline
       Pb+Pb @ 5020 GeV & 7.96 \\ \hline \hline
    \end{tabular}
    \label{tab:inelCrossX}
\end{table}
Once all the participant nucleons are determined, the binary collision positions are chosen at the centers of the wounded nucleon pairs. These binary collision positions are provided to the Pythia 8 model to sample the production point for hard processes.

\subsection{High energy partonic shower}

The hard scatterings are generated within Pythia 8~\cite{Sjostrand:2014zea}, using its multi-particle interactions (MPI) framework~\cite{Sjostrand:1987su} without initial and final state radiation.
Partonic showers are modeled using the newly developed initial-\matter{} (\imatter{}) and \matter{} generators for the initial state radiation (ISR) and final state radiation (FSR), respectively. The FSR using \matter{} has been described elsewhere, both outside the \jetscape/\xscape{} framework~\cite{Majumder:2013re,Cao:2017qpx}, and within~\cite{Cao:2017zih}. In this subsection, we describe the physics of the new \imatter{} model.

\begin{figure}[h!]
    \begin{center}
        \begin{tikzpicture}
            % \draw[gray] (-2,-1) grid (2,1);
            \draw[-latex, ultra thick, red!50!black] (0,0) -- (2,0) node[above] {$a$} node[midway, above] {$x$};
            \draw[-latex, ultra thick, blue!50!black] (0,0) -- (2,1) node[above] {$s$} node[midway, above, rotate=25] {$\frac{1-z}{z}x$};
            \draw[-latex, ultra thick, green!50!black]  (-2,0) node[above] {$p$} -- (0.1,0) node[midway, above] {$\frac{x}{z}$};
            \draw[-latex, thick] (-2,-0.5) -- (-0.5,-0.5) node[midway, above] {time};
        \end{tikzpicture}
    \end{center}
    \caption{A splitting process inside initial-state parton shower.}
    \label{fig:BcwrdSplt}
\end{figure}
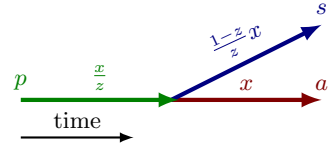

The initial state parton shower is given by a spacelike cascade where the last parton with the least negative virtuality ($-t_{\rm min}$) is the initial parton originating from the proton.
It is more efficient to sample the scattering first and then evolve the shower backward in time, starting from the hard scattering and evolving towards the incoming originating partons.
Backward evolution follows similar evolution equations to forward evolution.
However, since after each backward splitting, the parton `gains' energy $\left(x\to\frac{x}{z}\right)$ with $z < 1$, the phase space is limited using the parton distribution function (PDF).
The probability of one splitting is given by the vacuum splitting function $\hat{P}_{j\to i}(z)$ weighted by the PDF $f_j(x,t)$ as follows \cite{Ellis:1996mzs}
\begin{align}\label{eq:ProbSplit}
    \mathcal{P}_{p\to a}(z,t) = \frac{\alpha_s}{2\pi} \frac{\hat{P}_{p\to a}(z)}{z} f_p\left(x_p,t\right)\;,
\end{align}
This equation represents the probability of a parton $a$ with momentum fraction $x=\frac{p^+}{P_A}$ to originate in the split from a parton $p$ with momentum fraction $x_p=\frac{x}{z}$ as shown in Fig.~\ref{fig:BcwrdSplt}, with $P_A$ the incoming nucleon's momentum.

Since the PDF momentum fraction is limited to $\frac{x}{z} \leq 1$, the momentum fraction $z$ is bound from below by $z \geq x$. The splitting function's singularity is regulated due to this PDF's momentum fraction limit.

To sample the momentum fraction $z$ in the splitting function, as outlined in Eq.~\eqref{eq:ProbSplit}, one needs to determine the parton $a$'s virtuality $t$. This is done by sampling the Sudakov form factor, obtained by resumming the vacuum splittings, 
\begin{eqnarray}\label{eq:Sudakov}
    \Pi_i(t_1, t_2)
    = e^{ - \sum_p \int_{t_{1}}^{t_2} dt \frac{\alpha_s}{2\pi t} \int \frac{dz}{z}  \hat{P}_{p\to a}(z) \frac{f_p\left(x/z,t_1\right)}{f_i(x,t_2)}} ,
\end{eqnarray}
where $\Pi_i(t_1, t_2)$ represents the relative probability of backward evolution from parton $a$ to $p$ with their corresponding virtuality $-t_2$ to $-t_1$ without any additional splitting.
Starting with the parton $a$ with virtuality $-t_2$, we sample parton $p$'s virtuality  $t_1 \in [0, t_2]$ uniformly and use $\Pi_i(t_1, t_2)$ as the relative probability to decide whether this splitting happens or not.

In Monte Carlo simulations, the shower starts by generating the initial virtuality $-t_0$ of the initiating parton using the Sudakov form factor $\Pi_i(t_0,t_{\rm max})$.
The maximum modulus of the negative virtuality is limited by the transverse momentum of the scattering $\hat{p}_T$ as $t_{\rm max} = C_{v} \hat{p}_T^2$, where $C_v$ is a parameter between 0 and 1 and is typically set to $C_v \sim 0.25$ (as in the case of \matter{}) \cite{Majumder:2013re,Cao:2017qpx}.
%While this parameter can be dialed in a calibration with experimental data, such exercise has not been done in the current effort.
Subsequent splittings are generated by first sampling the virtuality $-t_p$ of the parent parton using the Sudakov form factor $\Pi_i(t_{p-1},t_p)$.
Then, the process is selected by sampling the splitting probability $\mathcal{P}_{j\to i}(z,t_p)$ and subsequently the momentum fraction $z$ is sampled. Once the light-cone momentum and virtuality of each parton in Fig.~\ref{fig:BcwrdSplt} is determined, transverse components can be determined by four-momentum conservation. The azimuthal angle of the split is random.

We employ the MPI functionality of Pythia to generate multiple hard scatterings.
Each hard scattering leads to two initial partons initiating independent ISR showers.
To ensure that the total energy available for the shower is limited to the energy of the proton, we start with the shower of the hardest splitting and rescale the momentum fraction $x$ of the subsequent showers by $x\to \frac{x}{1-\sum_i x_i}$, where $i$ runs over the generated showers.
After each splitting, we rotate all the partons in the shower such that the momentum of the parent parton is along the $z$-axis.

After generating the ISR of both initial partons undergoing a scattering $(\p_a,\p_b)$, the different rotations will lead to unaligned (non-collinear) 3-momenta $(\tilde{\p}_a,\tilde{\p}_b)$. An alternate way to think about this configuration is that two hard partons originate in two opposing, incoming nucleons. These are completely collinear with the protons. These then undergo multiple emission, leading to the formation of two hard virtual partons which will participate in the hard scattering, but are no longer collinear to each other.  
We compute the difference of momenta $\Delta \p = \p_a + \p_b - \tilde{\p}_a - \tilde{\p}_b$, which is added to the final state partons, immediately after hard scattering:
\begin{align}
    \tilde{\p}_c &= \p_c - \frac{\Delta \p}{2}\;,\quad
    \tilde{\p}_d = \p_d - \frac{\Delta \p}{2}\;.
\end{align}
The final state partons with momenta $(\tilde{\p}_c,\tilde{\p}_d)$ develop FSR with the \matter{} model~\cite{Majumder:2013re,Cao:2017qpx}.
Once the FSR complete, the final state of the hard partonic shower is hadronized following the Lund string model using Pythia, as described in Sec.~\ref{subsec:Fragmentation}.

The sampling of the ISR is performed similarly to Pythia, the main difference is how we perform the kinematic rotation that keeps the initial parton along the $z$-axis.
In Pythia, ISR is carried out simultaneously for both partons involved in the scattering.
Following each ISR radiation event, the system is boosted to the center of mass frame of the generated parton and the parton from the opposite side of the scattering.
Our approach involves simplifying the process by conducting the ISR for each parton individually.
This adjustment is made in anticipation of future enhancements, such as the incorporation of cold nuclear matter effects, where the partons may interact with nucleons before emitting radiation.
Consequently, and similar to the \matter{} generator, we also follow the space-time evolution of the showering partons. A Wigner-function based formalism is used to assign the production locations of the partons~\cite{Majumder:2013re}.  While these locations are not used in the current effort, they will play a role in the future case of interaction in cold nuclear matter prior to the hard scattering that produces jets. 

\subsection{Soft-hard correlations in particle production}

The \imatter{} model develops the ISR showers for partons $a$ and $b$, which undergo the hard scattering, and provide the \xscapegim{} simulator with their corresponding initiating parton\ $1$'s and $2$'s energy-momentum vectors $P_1^\mu$ and  $P_2^\mu$. 
This energy-momentum will have to be subtracted from the incoming nucleons in the \Glb{} model for the bulk sector particle production. 
The goal will be to designate one pair of constituent hot spots, one in each of the colliding nucleons, to have participated in the hard process. 

Inside each wounded nucleon, we sample the longitudinal momentum fractions for the three constituent quark hot spots from the valence quark PDF with a constraint that $\sum_{i=1}^3 x^q_i \le 1$~\cite{Shen:2022oyg}.\footnote{In this work, we use the CT10-NNLO PDF \cite{Gao:2013xoa} for protons and include nuclear modifications with the EPS09 for constituent quarks inside nucleons in heavy nuclei (Au or Pb). Varying the choices of the PDF and NPDF are not expected to qualitatively alter the conclusions of this study.}
The remaining energy and momentum of the incoming nucleon are assigned to a constituent gluon hot spot.
With the longitudinal momentum fraction $x_i^q$, we can compute the rapidity of the constituent quark hot spot $i$ at a given center of mass collision energy $\sqrt{s}$ as~\cite{Shen:2017bsr},
\begin{equation}
    y^q_i = \mathrm{arcsinh}\left(x^q_i\sqrt{\frac{s}{4 m_\mathrm{parton}^2} - 1} \right).
\end{equation}
The initial energy-momentum vector of the constituent quark hot spot $i$ is
\begin{equation}
    P_{q,i}^\mu = (m_\mathrm{parton}\cosh(y^q_i), 0, 0, m_\mathrm{parton}\sinh(y^q_i)).
\end{equation}
Here, we set $m_\mathrm{parton} = m_N/3$ for constituent quarks hot spots. Then the initial energy-momentum vector of the constituent gluon hot spot is
\begin{equation}
    P_{g}^\mu = P_N^\mu - \sum_{i = 1}^3 P_{q,i}^\mu.
\end{equation}
The spatial positions of the four constituent hot spots inside a nucleon are sampled from a 3D Gaussian profile with a width $B_G$ as in Eq.~\eqref{eq:nucleonProfile}.

%After Pythia 8 and \imatter{} models have determined the \textcolor{red}{originating parton energy-momenta that have to be subtracted from the} projectile and target nucleons, 
The \xscapegim{} simulator, taking input from the collision geometry from the \Glb{} model, determines the locations of the projectile and target nucleons.
%(this step is ignored for \pp{} and is only carried out on the nucleus side in \pPb{}).
Having determined the interacting nucleons, we randomly select a pair of constituent hot spots $(l, m)$ with their energy $P_{q/g, l}^0 > P_{1}^0$ and $P_{q/g, m}^0 > P_{2}^0$, respectively. Then, we subtract the hard energies and momenta from the constituent hot spots. This hot spot pair's remaining energies and momenta will be attributed to the remnant parts of their wounded nucleons. 
In cases where the energies of all hot spots are smaller than the hard originating parton's energy, we resample the triplet $\{x^q_i\}$ until at least one constituent hot spot carries enough energy.

Once the initial hard energy and momentum vectors $P_{1,2}^\mu$ are subtracted from the collision system, the \Glb{} model generates the initial soft strings from the other constituent hot spot pairs. These soft strings become energy-momentum source terms for the ensuing hydrodynamical evolution. For a detailed description of the soft string production, please refer to Refs.~\cite{Shen:2017bsr, Shen:2022oyg}.

After all the soft strings are produced, the remaining energy and momentum inside the wounded nucleon are treated as the energy-momentum vector for the nucleon remnant~\cite{Shen:2022oyg}, which is passed to the hadronization module to fragment together with the parton shower as described in Sec.~\ref{subsec:Fragmentation}. The nucleon remnants provide the balanced QCD color sources for the Pythia string fragmentation with the parton shower.

\subsection{Bulk dynamics and soft particle production}

Once the \Glb{} model produces the soft strings, they are treated as source terms for the system's energy-momentum tensor and net baryon current~\cite{Shen:2017bsr,Shen:2022oyg}, leading to the custodial equations, 
\begin{equation}
    \partial_\mu T^{\mu \nu} = J^\nu, \qquad 
    \partial_\mu J_B^\mu = \rho_B.
    \label{eq:hydroEOM}
\end{equation}
These hydrodynamic equations of motion are solved together with \textsc{neos-bqs}~\cite{Monnai:2019hkn}, a crossover equation of state (EOS) for the produced QCD matter. This EOS is constructed using recent lattice QCD calculations and extended to finite net baryon density with a Taylor expansion. It imposes the local strangeness neutrality condition $n_S\! = \!0$ and $n_Q\! =\! 0.4 n_B$ for net electric charge density.

In this work, we consider shear and bulk viscous effects during the hydrodynamic evolution. The shear stress tensor $\pi^{\mu\nu}$ and bulk viscous pressure $\Pi$ are evolved according to the Israel-Stewart type of equations up to the second order in spatial gradients~\cite{Israel:1979wp, Denicol:2012cn, Denicol:2018wdp}.

As the system evolves to the dilute region, we perform the Cooper-Frye particlization procedure to convert individual fluid cells on a hyper-surface with a constant energy density to hadrons~\cite{Cooper:1974mv, Shen:2014vra}. The Cornelius algorithm numerically determines the freeze-out hypersurface~\cite{Huovinen:2012is}. Excited resonance particles emitted from the kinetic freeze-out surface will decay to stable hadrons.

\subsection{Fragmentation of hard parton shower}
\label{subsec:Fragmentation}
Once the hard partonic shower reaches the non-perturbative virtuality scale $t_0$, the hard partons must be confined into hadrons.
Similar to the \jetscape{} framework, we will employ the Lund string fragmentation model within Pythia 8 to hadronize the hard partons.

The Lund string model employs the large $N$ (infinite) color limit, where the whole system is color neutral, with each parton carrying a color and/or an anti-color tag.
Strings connect color to anti-color tags, and these strings are stretched and broken to form hadrons.
Starting from the initial partons created at the end of the ISR, we generate new color tags for each radiation so that the color flow is exactly determined throughout the shower.

However, for each hard scattering from the MPI, one needs to generate two remnant partons going along the beam direction to complete a color neutral system.
The remnant partons are generated from the energy remaining from the \Glb{} model, which was not deposited in the hydrodynamic simulation.
The remnants, together with the final state partons of the hard parton shower, are then passed to another instance of Pythia 8, which hadronizes them to form hadrons.

In the subsequent section, we present the first results from this framework. We start with the simple case of no energy loss in either the initial or final state radiation and introduce energy loss in final state radiation in a later section. Consistency between the results of these sections will point to the small size of the produced medium.

\section{Results without final-state energy loss}
\label{sec:results}

In this section, we carry out numerical simulations with the \xscapegim{} simulator for \pp{} and \pPb{} collisions at $\snn = 5.02$~TeV. In the first set of results presented in this section, we  will not consider medium-induced energy loss from the interactions among the outgoing partons and the Quark-Gluon Plasma medium produced in these small systems (final-state energy-loss effects will be estimated in Sec.~\ref{sec:energy-loss}). The reasons for this setup are two-fold: The produced fireball in \pp{} and \pPb{} collisions is expected to be very small, and thus energy loss effects are expected to be minimal. Secondly, given the correlation between the energy momentum in the hard and soft sectors, one produces a different medium for a different set of hard partons (participating in hard scatterings). Thus, each hard scattering (Pythia) event requires a different (\Glb) initial state and a different 3+1D fluid dynamical (\MUSIC{}) simulation. As a result, these simulations become rather compute intensive. 

In the first subsection, we present a comparison with charged pions and jet spectra and $R_{pA}$ at $\snn = 5.02$~TeV. These data are used to fit our parameters (initially by eye). {The choice of the model parameters and their values are listed in Table~\ref{tab:ModelParams} in Appendix~\ref{App:ModelParams}}. Using these parameters, we then explore the correlation between the presence of hard scattering and transverse energy at both forward and midrapidity in Sec.~\ref{sec:soft-hard}. In the final subsection, we carry out a small-scale (limited training design with low statistics simulations and a constrained parameter set) Bayesian analysis to allow for an exploration of the parameter space of our model. 

\subsection{Hadron and jet spectra in \pp{} and \pPb{} collisions}

\begin{figure}[h!]
    \centering
    \includegraphics[width=0.45\textwidth]{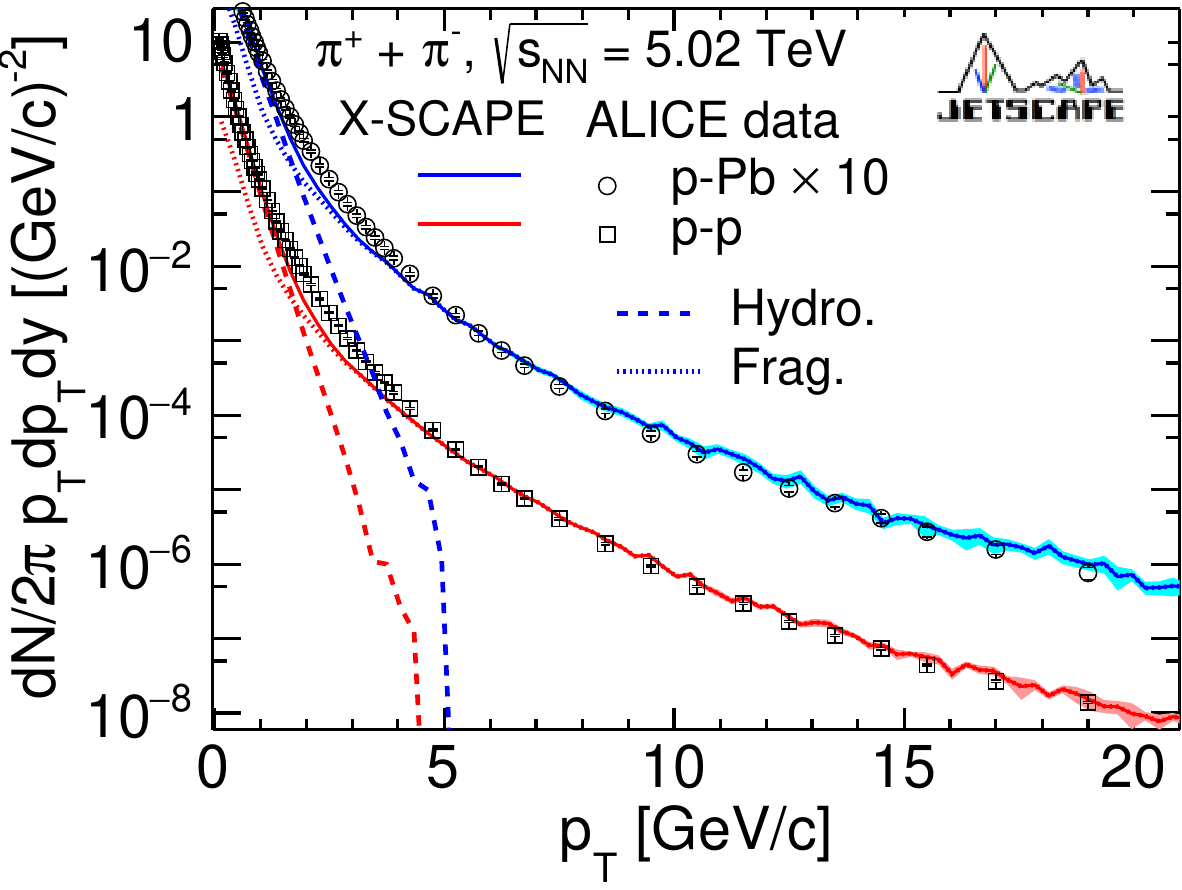}
    \caption{Transverse momentum spectra of charged pions in minimum bias \pp{} and \pPb{} collisions at $\snn = 5.02$~TeV. Separate contributions to the hadron spectra from the bulk sector and the hard sector are labeled as ``Hydro.'' and ``Frag.'', respectively. The ALICE data are from Refs.~\cite{ALICE:2016dei,ALICE:2019hno}. }
\label{fig:hadronspctra}
\end{figure}

Figure \ref{fig:hadronspctra} shows the transverse momentum ($p_T$) spectra of charged pions in minimum bias \pp{} and \pPb{} collisions at $\snn = 5.02$~TeV compared with the ALICE measurements. The full \xscapegim{} results reasonably describe the particle production for $p_T$ from 0 up to 20 GeV in small systems. Looking at contributions from individual components, the spectra of charged pions are dominated by the soft particle production from the hydrodynamic phase for $p_T < 2$~GeV. The fragmentation from high-energy partons contributes $\sim 15\%$ to the particle yield in this $p_T$ region. This result is consistent with those from the dynamical core-corona model~\cite{Kanakubo:2022ual,Werner:2007bf,Aichelin:2008mi}.
The particle production with $p_T > 4$ GeV is dominated by the fragmentation of hard partons.
In our simulations, the switching $p_T$ scale from soft to hard particle production is controlled by the model parameter $\hat{p}_{T,\mathrm{min}}$, which represents the lower limit of hard scattering in the MPI. 

\begin{figure}[h!]
    \centering
    \includegraphics[width=0.45\textwidth]{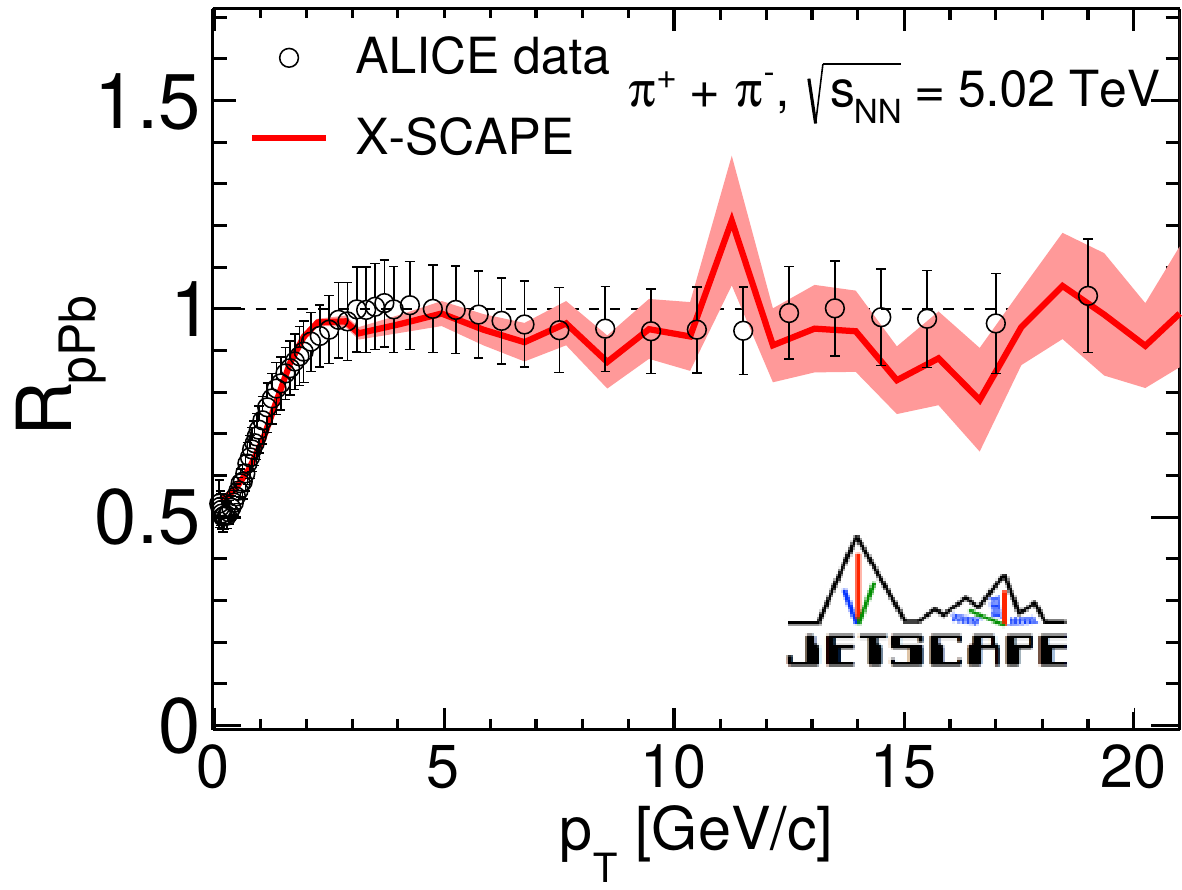}
    \caption{The nuclear modification factor $R_{\rm pPb}$ as a function of transverse momentum for charged pions in minimum bias \pPb{} collisions at $\snn = 5.02$~TeV (ALICE data from \cite{ALICE:2016dei}). }
\label{fig:RpPbhadron}
\end{figure}

It is noteworthy that there is some tension in reproducing the $p_T$-spectra at the intermediate range ($2 < p_T < 5$ GeV/c). This discrepancy might arise from a lack of quark coalescence processes between the shower and medium partons during hadronization, which has a significant contribution in this $p_T$ region \cite{Greco:2003xt,Fries:2003vb,Zhao:2020wcd}. At this time, a fully developed, calibrated module that carries out \emph{re-combinatoric} hadronization, incorporating both the hard and bulk sectors, within the \xscapegim{} simulator, is not yet available.  

Taking the ratio between the charged pion spectra in \pPb{} and \pp{} collisions, we present the nuclear modification factor $R_{\rm pPb}$ in minimum bias \pPb{} collisions at 5.02 TeV in Fig. \ref{fig:RpPbhadron}. The \xscapegim{} results provide a quantitative description of the $R_{\rm pPb}$ across the full $p_T$ range.
For transverse momenta $p_T > 4$ GeV, both the \xscapegim{} calculations and ALICE measurements are consistent with unity within uncertainties, suggesting negligible final-state energy loss for the shower partons. In the \xscapegim{} calculations, the value of $R_{\rm pPb}$ goes below 1 for $p_T < 4$ GeV because the dominant particle production mechanism transits from perturbative hard scatterings to non-perturbative soft production, in which particle yields scale slower than the number of binary collisions.

We note that the observation of $R_{\rm pPb} < 1$ at $p_T \lesssim 4$ is sometimes interpreted as arising from initial-state nuclear shadowing effects in the $Pb$ nucleus. Our model provides an alternative interpretation that soft particle production from a fluid dynamic picture can produce the suppression of $R_{\rm pPb}$ at low $p_T$ as well. 
The current implementation of the \xscapegim{} simulator neither includes initial-state nuclear shadowing nor the intrinsic transverse momentum broadening resulting from initial multiple parton scattering (Cronin effect)~\cite{Cronin:1973fd,Cronin:1974zm} in \pA{} collisions. The future, planned inclusion of these effects in \imatter{} may become crucial for a comprehensive understanding of $R_{\rm pPb}$, at intermediate $p_T$ in \pA{} collisions~\cite{Cronin:1973fd,Wang:1998ww}.

\begin{figure}[h!]
    \centering
    \includegraphics[width=0.45\textwidth]{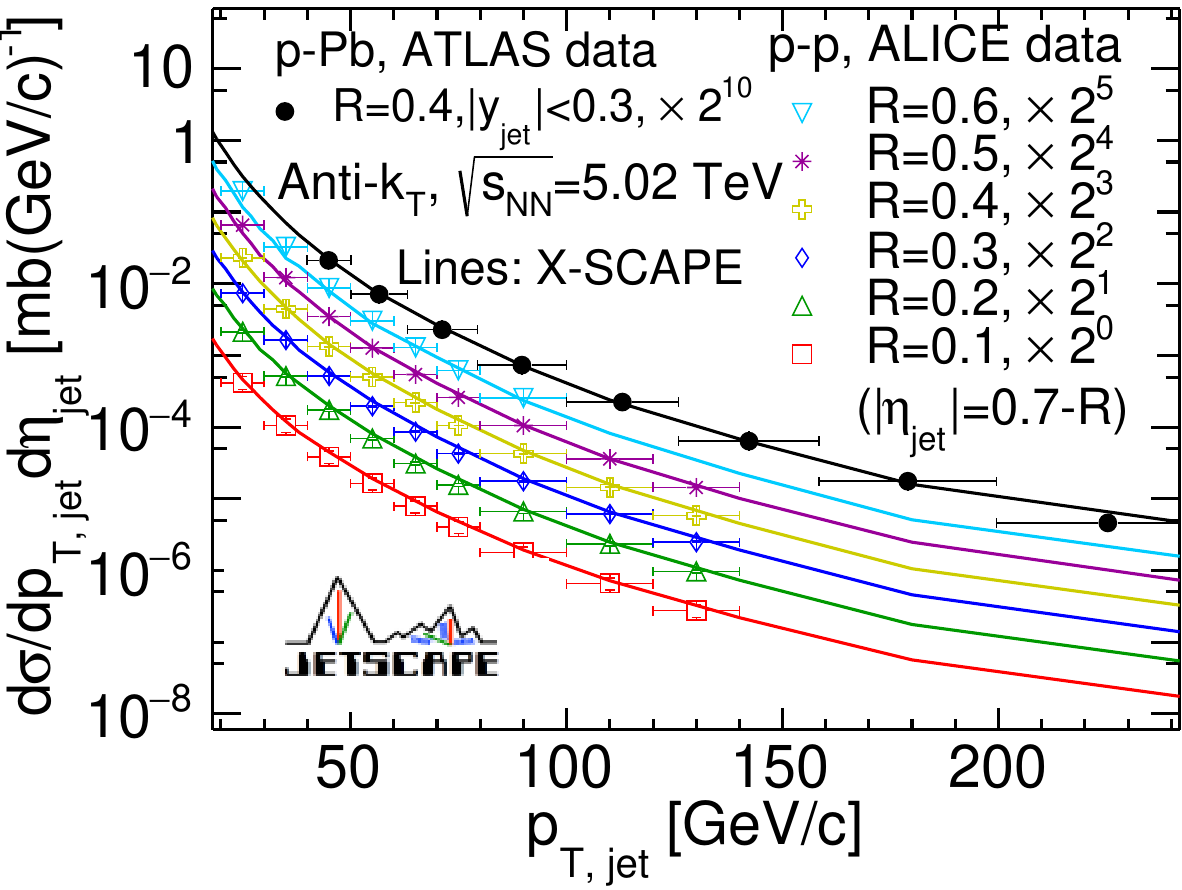}
    \caption{The full jet $p_T$-differential cross sections at $\snn = 5.02$ TeV for \pp{} collisions with jet cone of radius $R = 0.1 - 0.6$ and 0-90\% \pPb{} collisions with $R = 0.4$. The ALICE and ATLAS measurements are from \cite{ALICE:2019qyj,ATLAS:2014cpa}.}
\label{fig:jetspcetra}
\end{figure}
\begin{figure}[h!]
    \centering
    \includegraphics[width=0.45\textwidth]{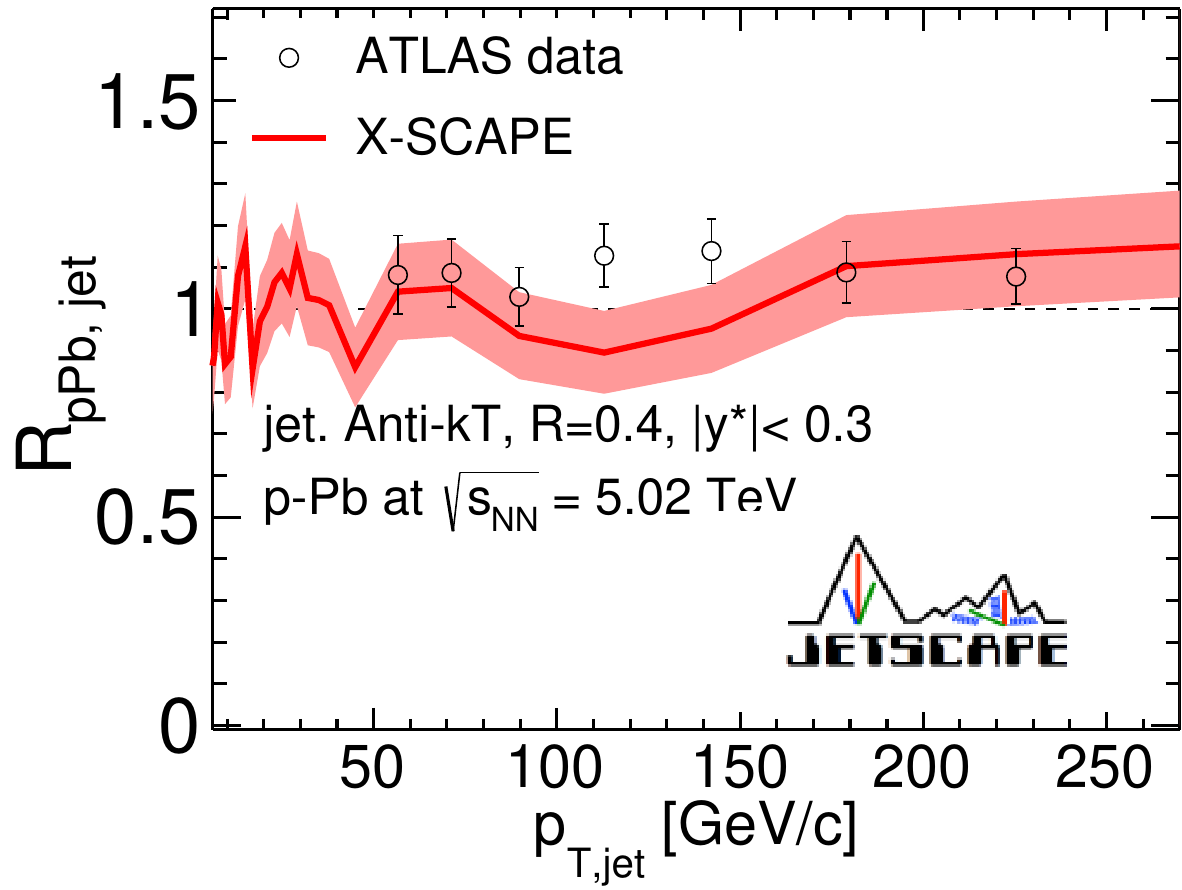}
    \caption{The nuclear modification factor $R_{\rm pPb, jet}$ as a function of transverse momentum for the full jets in 0-90\% \pPb{} collisions at $\snn =  5.02$~TeV. The ATLAS data is from \cite{ATLAS:2014cpa}.}
\label{fig:RpPbjet}
\end{figure}

Figure \ref{fig:jetspcetra} shows the inclusive full jet $p_T$-differential cross section in \pp{} collisions for the pseudo-rapidity ranges $\vert\eta_{\rm jet}\vert < 0.7 -R$ with jet resolution parameters $R = 0.1, 0.2, 0.3, 0.4, 0.5, 0.6$. The full jet $p_T$-differential cross section in 0-90\% \pPb{} collisions is also shown in Fig.~\ref{fig:jetspcetra} for $R = 0.4$ with jet rapidity $\vert y^*_{\rm jet}\vert < 0.3$ in the center-of-mass frame. Jets are reconstructed using the anti-$k_T$ algorithm implemented in FastJet \cite{Cacciari:2011ma}.
The results from the \xscapegim{} simulator agree well with the ALICE and ATLAS data for all cases. 
Our analysis assumes a perfect background subtraction of the underlying event activity from the soft medium and only clusters partons from the shower off a hard parton emanating from one of the MPI scatterings.
The successful description of jet spectra, in \pp{} collisions, establishes a reliable baseline for the study of nuclear modification in \pA{} collisions.

The nuclear modification factor $R_{\rm pPb, jet}$, as a function of the transverse momentum of the jet, for 0-90\% \pPb{} collisions, at $\snn = 5.02$~TeV, is presented in Fig.~\ref{fig:RpPbjet}. The analysis is performed at mid-rapidity ($\vert y^* \vert < 0.3$).
At mid-rapidity, the observed $R_{\rm pPb, jet}$ is consistent with unity within the systematic uncertainties. It reveals no significant modification of the total yield of jets relative to the nuclear geometric expectation in 0-90\% \pPb{} collisions. The \xscapegim{} calculation here, without incorporating the effects of energy loss by the initial hard partons, successfully reproduces the experimental results. This result aligns with the behavior observed in the minimum bias $R_{\rm pPb}$ of inclusive charged $\pi^{\pm}$ in Fig. \ref{fig:RpPbhadron}, where  no discernible nuclear modifications are evident in \pPb{} collisions.

Overall, the \xscapegim{} results presented in this subsection benchmark the particle production in \pp{} and \pPb{} collisions without medium-induced energy loss effects. They provide consistent descriptions of the data for both jet and hadron production in \pPb{} collisions. For the case of hadron spectra, contributions from bulk dynamics included a fluid dynamical simulation followed by Cooper-Frye hadronization, while contributions from the hard sector included the fragmentation of strings stretched between the hard high-$p_T$ showers and beam remnants. The \xscapegim{} simulator provides a simulation where there is energy-momentum conservation between these two sectors, event-by-event. These combined together to produce the first seamless description of hadron production at all $p_T$ ($0\leq p_T \leq 20$~GeV) in Fig.~\ref{fig:RpPbhadron}. The effects of this energy balance between the hard and bulk sectors will be studied briefly in the next section.

\subsection{Soft-hard correlations in \pp{} and \pPb{} collisions}
\label{sec:soft-hard}

In this subsection, a simple study of the soft-hard correlation from global energy-momentum conservation in small systems is presented. Subtracting the energies and momenta of hard partons, from the bulk initial conditions, generates event-by-event anti-correlations between particle production from bulk and hard sectors. In this subsection, we briefly study these correlations. 

\begin{figure}[h!]
    \centering
    \includegraphics[width=0.45\textwidth]{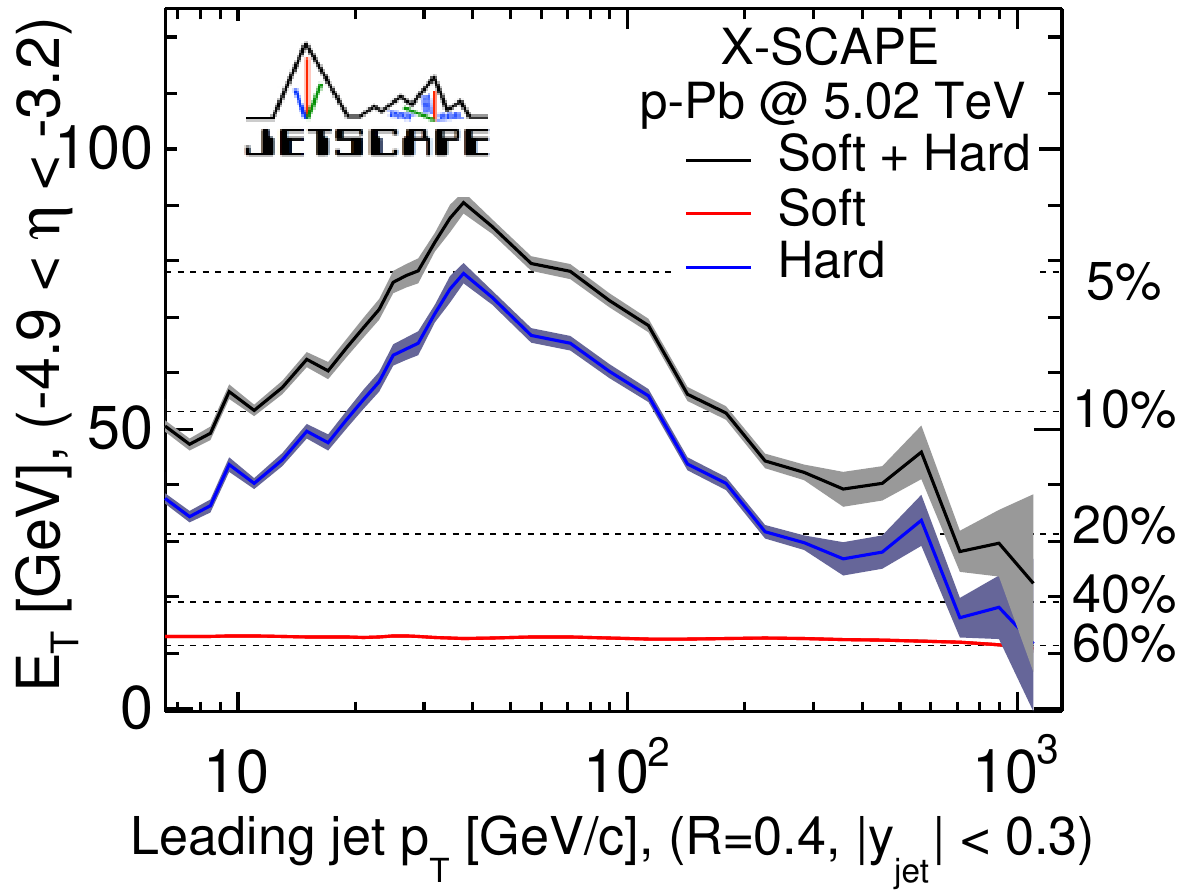}
    \caption{The transverse energy of hadrons produced from the bulk sector (red line), fragmentation hadrons from the hard sector (blue line) and total (black line) at backward rapidity range ($-4.9<\eta<-3.2$) as a function of leading jet $p_T$ in \pPb{} collisions at $\snn=5.02$~TeV. The dashed lines indicate the boundaries for centrality bins.}
\label{fig:pPbforward}
\end{figure}

No fits to experimental measurements on $p_T$ versus event activity, as measured by ALICE (at lower $p_T$), or centrality dependent spectra, as measured by ATLAS and CMS (at higher $p_T$), are included. These will require an extensive Bayesian analysis to fully calibrate the model (either with or without correlation with event activity), followed by high statistics simulations to see the effect of jet $p_T$ on separate event activity bins. Our goal in this effort, is to demonstrate that the \xscapegim{} soft-hard model with exact energy-momentum conservation, does indeed contain non-trivial correlations between the hard energy of a produced jet or hadron, and the event activity measured at various rapidities.

{{
Studying soft-hard correlation requires a large number of event-by-event simulations. 
To speed up simulations in the bulk sector, we have trained a deep neural network to emulate (3+1)D hydrodynamic simulations and to map initial-state longitudinal energy profiles ($d E/d\eta_s$) to final-state transverse energy distributions ($dE_T/d\eta$), event by event. This neural network was trained with 10,000 real hydrodynamic simulations.
We tested the trained neural network and found that the emulator predicted $dE_T/d\eta$ was tightly correlated with the validation results from the real model, which is sufficient for us to study the \emph{qualitative} features of soft-hard correlation in our model. For future quantitative comparisons with experimental measurements, we will perform systematic tests to determine whether the machine learning routine can cover the full fluctuation spectrum of the real model with sufficient accuracy.
Nevertheless, with the speed up enabled by this neural network, we can efficiently (and perhaps approximately) predict the final state transverse energy distributions from the bulk sector based on the initial state $dE/d\eta_s$ after subtracting the energy generated by the hard process and study the soft-hard correlation in this subsection.
}}

Figure \ref{fig:pPbforward} presents the collision event's transverse energy ($E_T$) in the backward lead-going side ($-4.9 < \eta < -3.2$) as a function of the leading jet $p_T$ at mid-rapidity in \pPb{} collisions at $\snn = 5.02$~TeV.
The red line shows the transverse energy from bulk medium hadrons. This soft component of the $E_T$ exhibits a mild decreasing trend with increasing leading jet $p_T$, because of a larger energy reduction in the \Glb{} initial conditions when the collision event produces a jet with higher energy.

The blue line shows a significant non-monotonic leading jet $p_T$ dependence of the backward transverse energy from the hard parton shower. The magnitude of $E_T$ increases with the leading jet $p_T$ and reaches the maximum at around $p^\mathrm{jet}_{T,\mathrm{peak}} \approx 50$~GeV in \pPb{} collisions. This dynamically generated momentum scale results from the competition between the energy of the jet shower at mid-rapidity and the energies of corresponding nucleon remnants connected with the hard partons. 

For jets with $p_T \lesssim 50$ GeV, the typical nucleon remnants carry larger momentum than those in the outgoing parton shower. So, the fragmented hadrons are distributed over a wider range of rapidity. In this kinematic region, the transverse energy $E_T$ in the backward rapidity positively correlates with the leading jet $p_T$ at mid-rapidity because higher energy jets radiate more and produce more particles, many of which form parts of the string stretching with the remnants. For jets with $p_T \gtrsim 50$ GeV, the outgoing parton shower with ever larger transverse momentum starts to win over the nucleon remnants and pulls more fragmented hadrons towards midrapidity. It results in a suppression of the $E_T$ in the backward rapidity. Based on this mechanism, we expect this momentum scale $p^\mathrm{jet}_{T,\mathrm{peak}}$ to be an increasing function of the collision energy $\snn$.

The total transverse energy (as a function of the $p_T$ of the leading jet) exhibits a similar behavior to transverse energy from the hard sector, as this portion dominates in jet-triggered events. This non-monotonic leading jet $p_T$ dependence of the transverse energy is essential to understanding the correlation between event activity and jet production in small systems. The LHC experiments usually use the backward transverse energy or particle multiplicity to define the event centrality~\cite{ATLAS:2014cpa,ALICE:2014xsp,ALICE:2017svf}. We show the ATLAS centrality definition as dashed lines in Fig.~\ref{fig:pPbforward}. When one triggers jets or hadrons with $p_T \lesssim 50$~GeV, the positive correlation between $E_T$ and leading jet $p_T$ results in a bias to select more central collisions with a high jet $p_T$ trigger. When the trigger energy is larger than the $p^\mathrm{jet}_{T,\mathrm{peak}}$, the opposite $p_T$ dependence of $E_T$ correlates collision events with less event activity with the higher jet $p_T$ trigger.

Interestingly, these two distinct correlations between event activity and hard triggers have been observed in the ALICE~\cite{ALICE:2017svf} and ATLAS experiments~\cite{ATLAS:2014cpa}. The \xscapegim{} results in Fig.~\ref{fig:pPbforward} shows qualitative agreements with those observations. The ALICE measurements showed that the high-$p_T$ hadron triggers around 10~GeV generated a selection bias towards collisions with more event activity in the backward rapidity region, in line with the \xscapegim{} results for the leading jet $p_T \lesssim p^\mathrm{jet}_{T,\mathrm{peak}}$. The ATLAS measurements showed an $R_{{\rm pPb, jet}} > 1$ for $p_T \gtrsim 200$ GeV in peripheral \pPb{} collisions while $R_{{\rm pPb, jet}} < 1$ for central \pPb{} collisions. This measurement reflects the negative correlation between $E_T$ and leading jet $p_T$ for $p_T \gtrsim p^\mathrm{jet}_{T,\mathrm{peak}}$ in Fig.~\ref{fig:pPbforward}. Quantitative comparisons with these measurements require high statistics calculations to performed post a large scale Bayesian analysis, which will be pursued in future publications.

\begin{figure}[h!]
    \centering
    \includegraphics[width=0.45\textwidth]{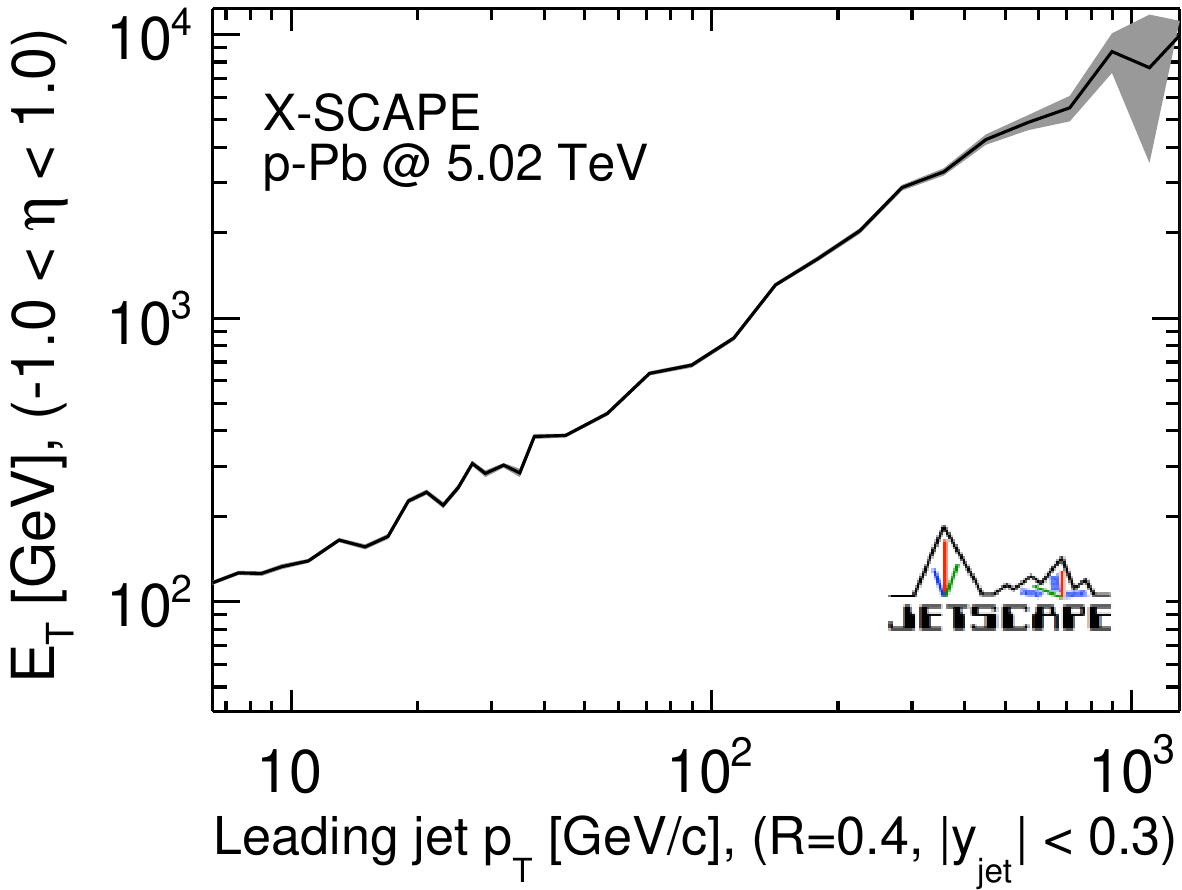}
    \caption{The total transverse energy at middle rapidity range ($-1<\eta<1$) as a function of leading jet $p_T$ in \pPb{} collisions at $\snn=5.02$~TeV.}
\label{fig:pPbmiddle}
\end{figure}

In Figure \ref{fig:pPbmiddle}, we present the total transverse energy at mid-rapidity, as a function of the leading jet transverse momentum in \pPb{} collisions. As anticipated, the total $E_T$ at middle rapidity exhibits a monotonic increase with the leading jet $p_T$, predominantly driven by the contributions from hard processes.
In the proton-going side, we find the forward $E_T$ has a qualitatively same correlation as that in backward ($Pb$-going) rapidity region.

\begin{figure}[h!]
    \centering
    \includegraphics[width=0.45\textwidth]{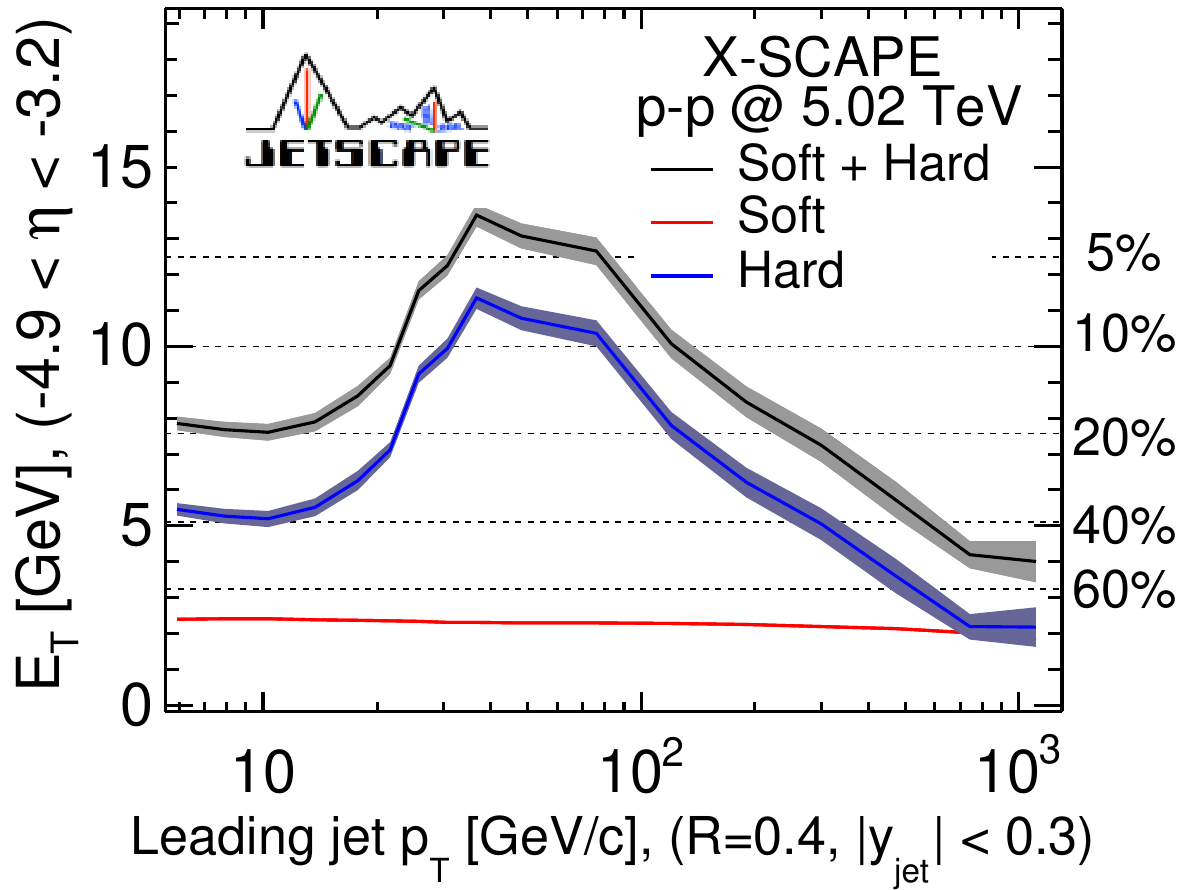}
    \caption{Similar to Fig. \ref{fig:pPbforward}, but for \pp{} collisions at $\snn = 5.02$~TeV.}
\label{fig:ppforward}
\end{figure}

\begin{figure}[h!]
    \centering
    \includegraphics[width=0.45\textwidth]{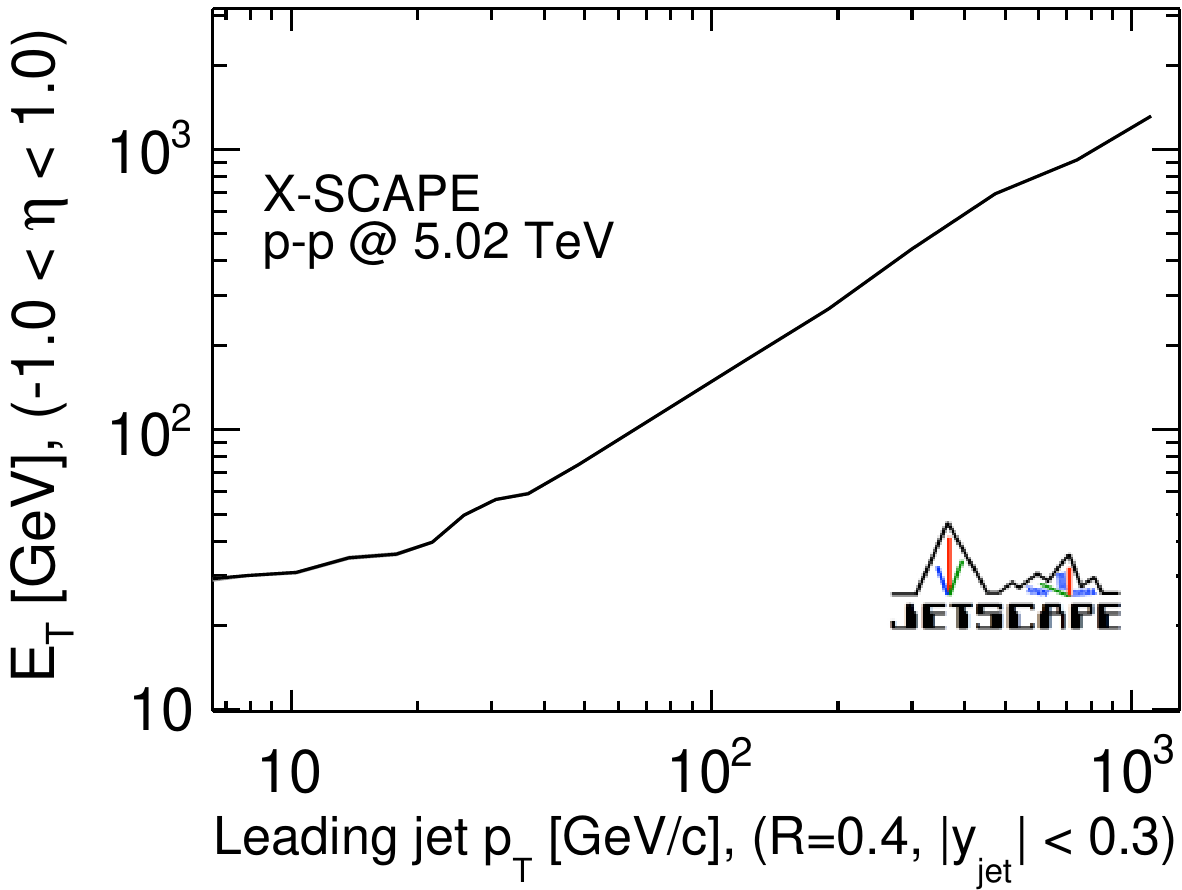}
    \caption{Similar to Fig. \ref{fig:pPbmiddle}, but for \pp{} collisions at $\snn = 5.02$~TeV.}
\label{fig:ppmiddle}
\end{figure}

Similar correlations between the collision event $E_T$, in forward and mid-rapidity, and the leading jet $p_T$ are present in \pp{} collisions, as shown in Figs.~\ref{fig:ppforward} and \ref{fig:ppmiddle}. Notably, the $E_T$ at the forward rapidity displays similar non-monotonic behavior with respect to the leading jet $p_T$ as in \pPb{} collisions. The transition momentum scale $p^\mathrm{jet}_{T,\mathrm{peak}}$ seems to be around 50~GeV, independent of the collision system. 

Studying the correlation between event activity and high $p_T$ triggers in different collision systems will be valuable in verifying the underlying physics picture presented here. It will also be interesting to study the collision energy dependence of this dynamically generated momentum scale. These will be addressed in upcoming efforts. 

\subsection{Parameter space exploration with Bayesian analysis}
\label{sec:Bayesian}

In this subsection, we would like to explore the model parameter space of the \xscapegim{} simulator and individual parameters' effects on final-state observables. In this particular simulator, designed within the \xscapegim{} simulator, we define a 15-dimensional parameter space, including variables that control particle production in soft and hard sectors. The details about model parameters are explained in the Appendix~\ref{App:ModelParams}. 

Because simulating event-by-event \pp{} and \pA{} collisions with the multi-stage \xscapegim{} simulator is computationally demanding, we need to train model emulators for the interested experimental observables and explore the model parameter space efficiently. With limited computational resources, we perform numerical simulations at 80 training points in a Latin-Hypercube design. We focus on the identified particle (pions, kaons and protons) $p_T$-differential spectra with $p_T < 20$~GeV in minimum bias \pp{} collisions at $\snn = 5.02$~TeV.

\begin{figure}[t!]
    \centering
    \includegraphics[width=0.45\textwidth]{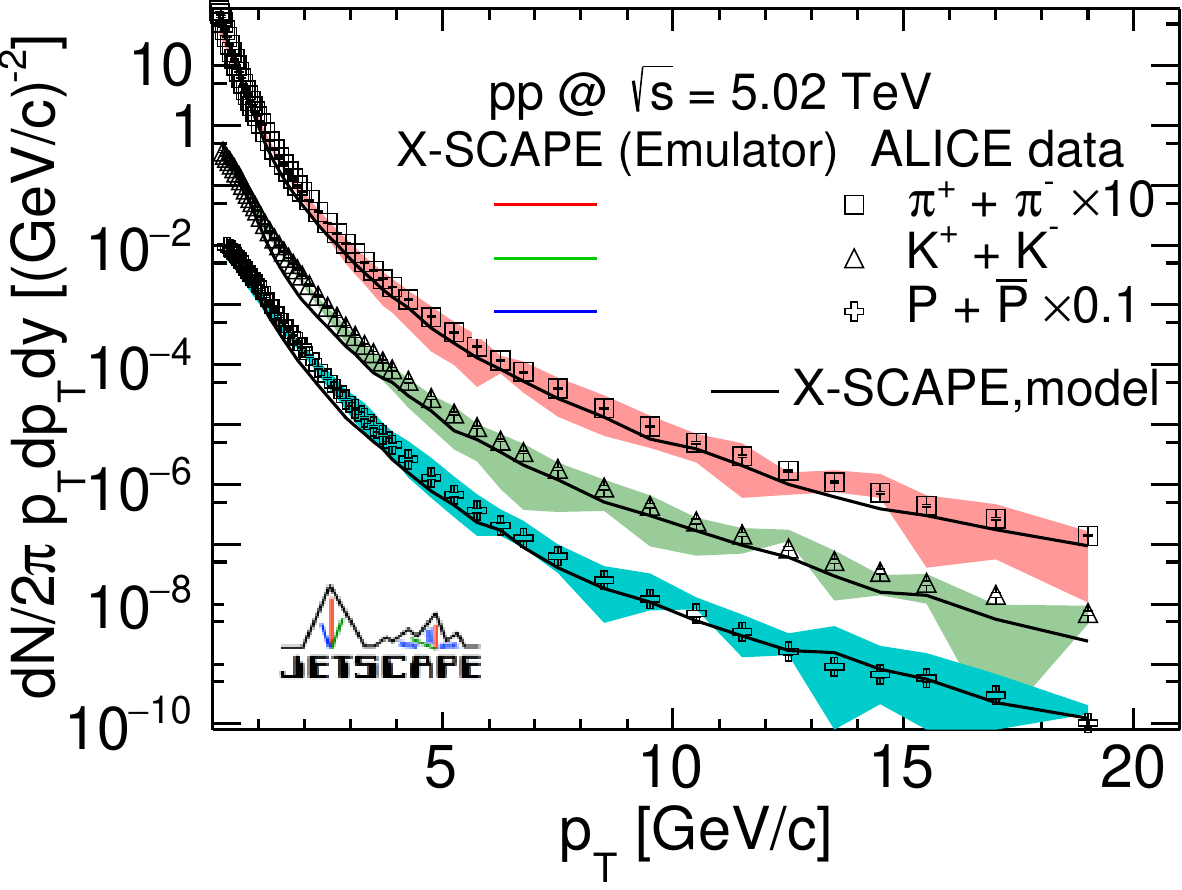}
    \caption{Transverse momentum spectra of pions, kaons, and protons. The bands present the model emulator prediction from 200 random parameter set samples drawn from the posterior distribution. The black lines are the real model calculation using the median values of the parameters from the Bayesian calibration. The ALICE data is from \cite{ALICE:2016dei}.}
\label{fig:pp_pikp_Bayesian}
\end{figure}

With the trained model emulator, we perform Markov Chain Monte Carlo (MCMC) in the 15-dimensional parameter space by constraining the model results with the ALICE measurements. The posterior distributions of the model parameters are shown in Appendix~\ref{App:ModelParams}. As shown in Fig.~\ref{fig:pp_pikp_Bayesian}, the \xscapegim{} simulator provides a reasonable description of the identified particle spectra in minimum bias \pp{} collisions. Although the model emulators were trained with a limited training design, we validate our Bayesian calibration with the real model calculation using the median values of the model parameters from the posterior distribution. The real model calculations (shown in the black solid line) are consistent with the predictions from model emulators. 

This exploratory study presented here paves the way for a more systematic Bayesian inference analysis in small systems with the \xscapegim{} simulator, in an upcoming work. Such an analysis will incorporate more parameters within the model and include a larger data set, beyond hadrons and jets.

\section{Estimate of energy loss}
\label{sec:energy-loss}

In the preceding section, we presented results for the spectra and nuclear modification factors for charged hadrons and jets, using the \xscapegim{} simulator (\Glb{} + pythia + \imatter{} + \MUSIC{} + \matter{}). For both \pp~and $p$-$Pb$ collisions, the hard partonic showers were calculated in the vacuum, ignoring any jet-medium interactions. This was carried out assuming that the medium formed within \pp{} and \pPb{} collisions would be too small to induce any energy loss. This assumption worked well in comparisons between the simulation results and experimental data (see Figs.~\ref{fig:hadronspctra}--\ref{fig:RpPbjet}).

\begin{figure}[h!]
    \includegraphics[width=0.31\textwidth]{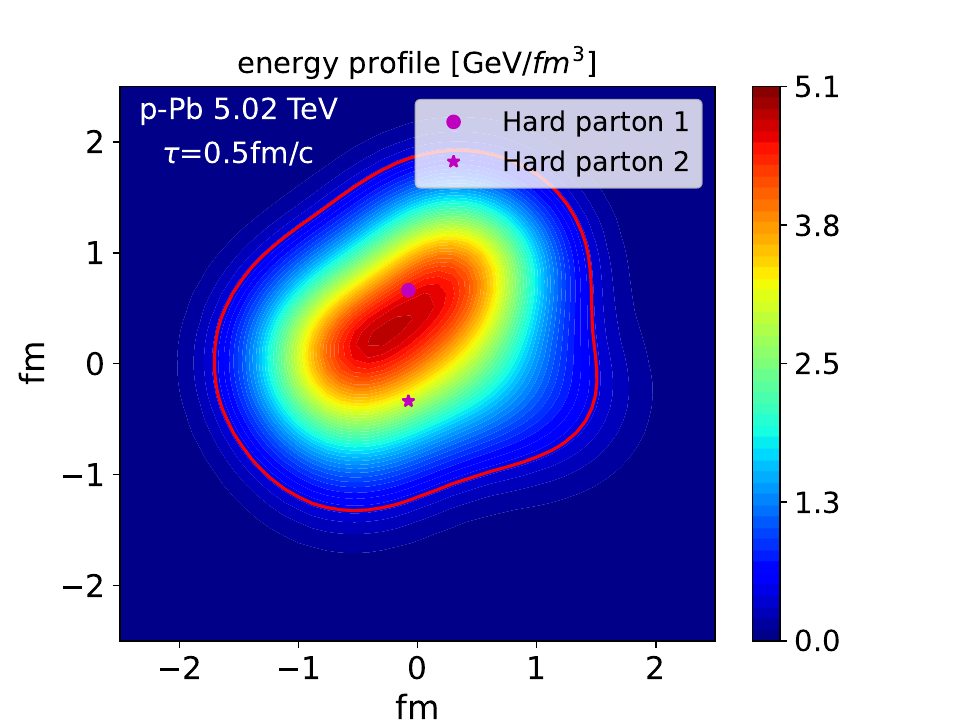} \includegraphics[width=0.31\textwidth]{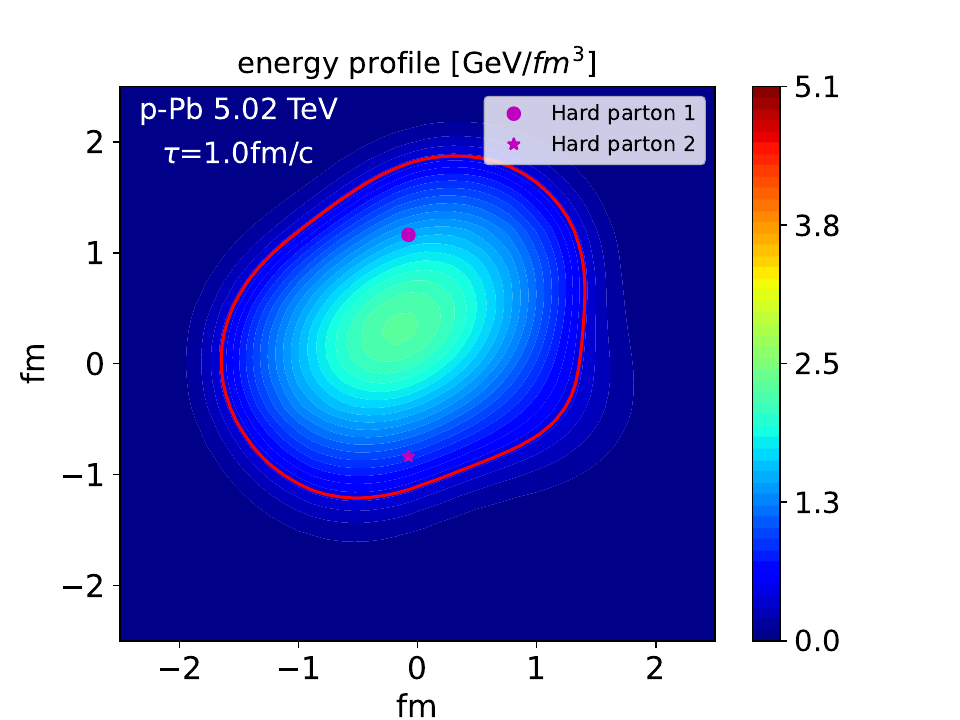} \includegraphics[width=0.31\textwidth]{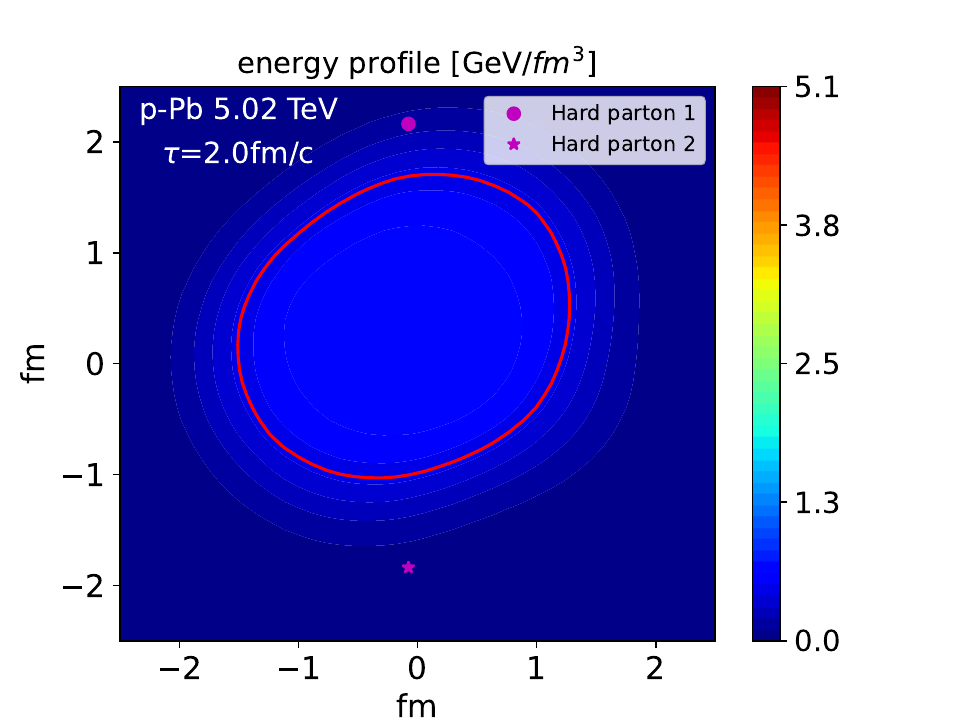}
    \caption{The bulk energy density and the hard partons position time evolution in the transverse plane at midrapidiy. The hard partons (purple dots) are moving in $\pm y$ directions with the speed of light. The red contour is the freeze-out line for hydrodynamics. The 3 snapshots are at $\tau=0.5, 1$ and $2$~fm/c. }
    \label{fig:energy_profile}
\end{figure}

\begin{figure}[h!]
    \includegraphics[width=0.45\textwidth]{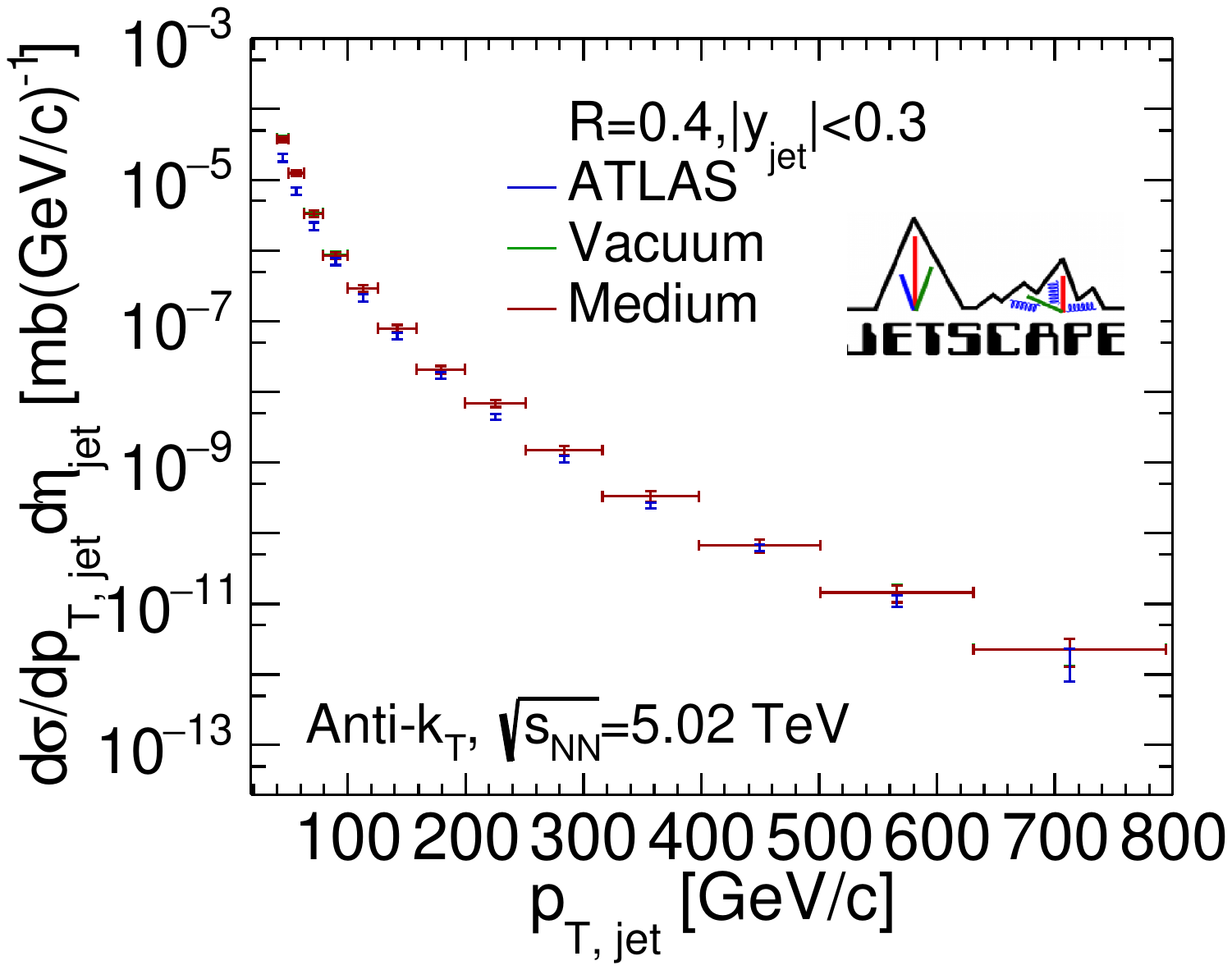}
    \includegraphics[width=0.45\textwidth]{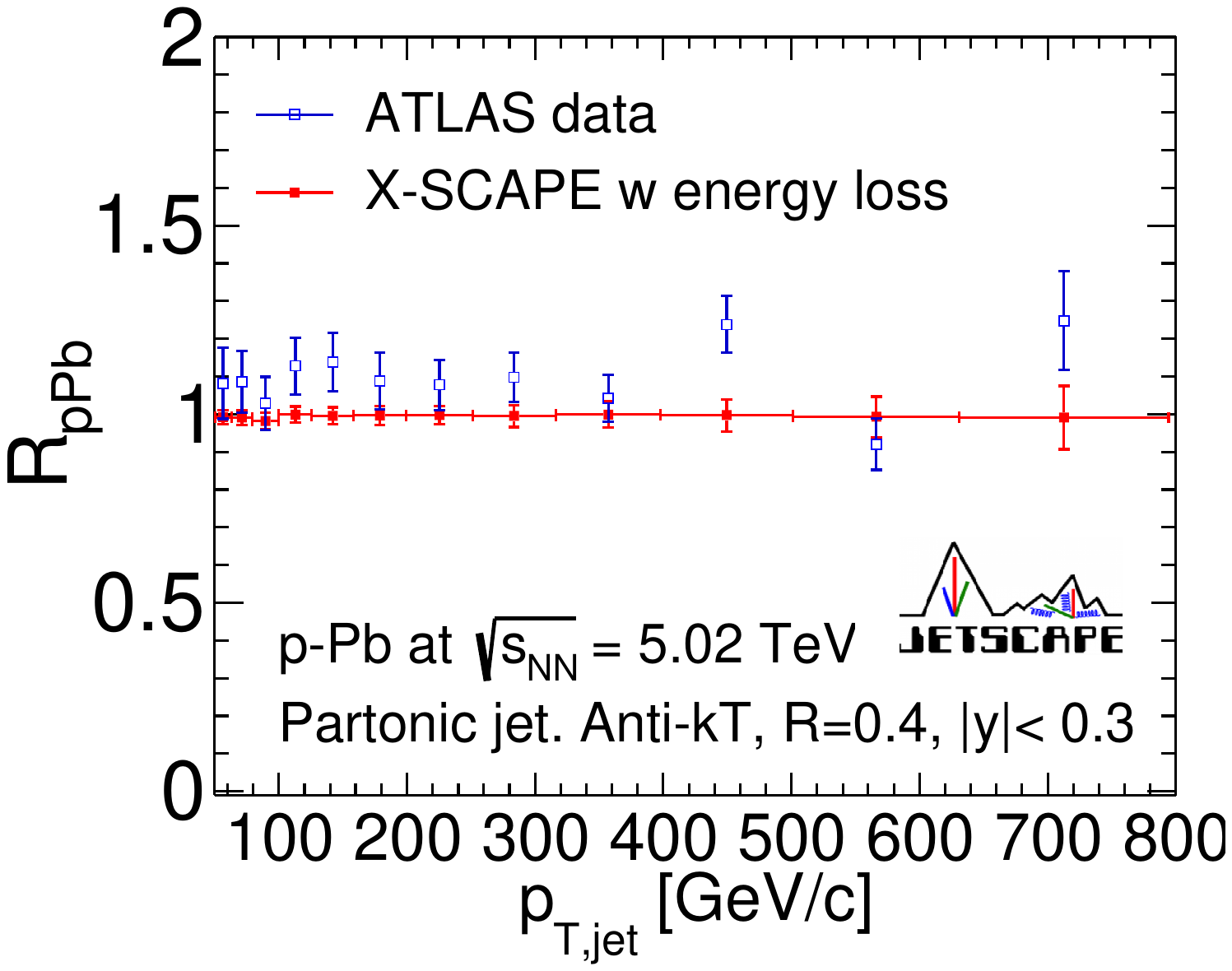}
    \caption{
        Partonic jet spectrum for \pp\ and $p$-$Pb$ collisions using, using anti-$k_T$ algorithm with $R=0.4$.
        On the top panel vacuum (red) and medium-modified (green) showers are shown with boxes.
        The bottom panel shows the nuclear modification factor $R_{\rm pPb}$.
        Compared with the ATLAS data \cite{ATLAS:2014cpa}.
    }
    \label{fig:PartonJetRpPb}
\end{figure}

\begin{figure}[h!]
    \includegraphics[width=0.45\textwidth]{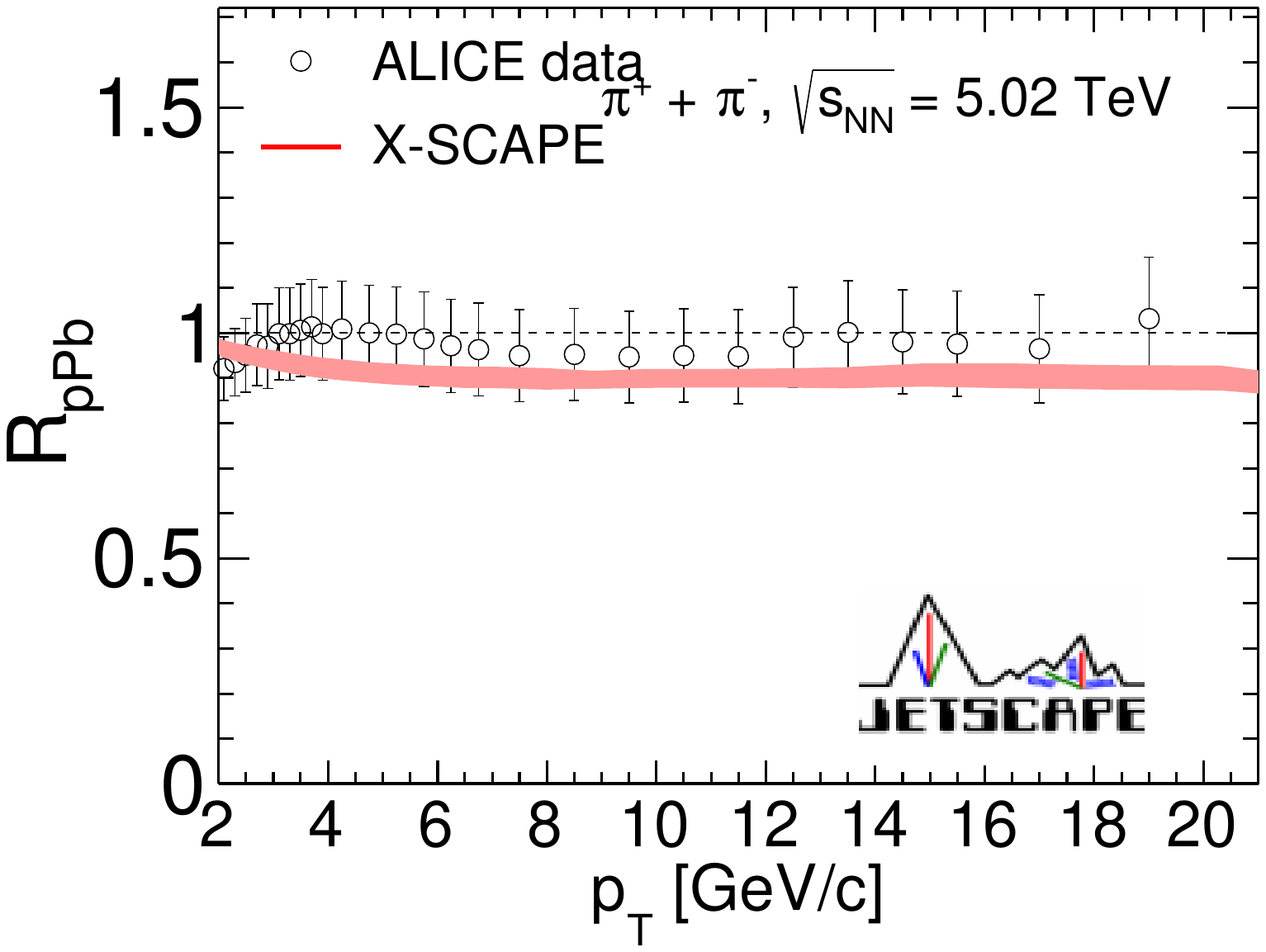}
    \caption{
        The nuclear modification factor $R_{\rm pPb}$ for charged pions calculated solely from the hard sector, compared with the ALICE data~\cite{ALICE:2016dei}.
    }
    \label{fig:PionRpPb}
\end{figure}

This section will explore the effect of allowing for medium-induced energy loss in the parton showers produced in the small systems.
The reason for not presenting these results first is that the energy and momentum deposited in \Glb{}, which seeds the evolving soft medium, depends on the amount of energy-momentum delegated to the hard processes. Thus, the expanding QGP is different in each event. As a result, one has to carry out a 3+1D fluid dynamical simulation in each event.  
This is extremely computationally expensive.
To compensate for this, we simulate the hard scattering and partonic shower in two steps:
\begin{enumerate}
    \item Hard scattering and initial state shower:
        We run $N_{\rm scattering}$ events, where we use Pythia to generate the hard scattering and I-\matter{} to simulate the initial state shower to obtain the initial partons.
    \item Final state shower and energy loss:
        For each event from the previous step, we subtract the energy from the \Glb{} initial state, which is then used to seed the hydrodynamic evolution.
        The hydrodynamic profile is saved and can then be used to run $N_{\rm shower}$ \matter{} shower events, now including medium-induced energy loss.
\end{enumerate}
Using this approach, we enhance the statistics of the energy loss portion rather than the full initial hard scattering cross section.
Given that the initial state of the \matter{} shower is saved to disc, we can use the same initial state but without including the energy loss to obtain a baseline vacuum evolution to compare against. Calculating the nuclear modification factor in this manner, removes the statistical error of the initial state from the process. Note, our goal in this section is to estimate the shift in $R_{\rm pPb}$ from energy-loss as accurately as possible. We are not trying to produce fully realistic results from the simulator will all aspects of statistical error included.  

In prior efforts using the \matter{} generator in the \jetscape{} framework, such as in Refs.~\cite{JETSCAPE:2022jer,JETSCAPE:2022hcb}, simulations included jet coherence effects~\cite{Kumar:2019uvu}, which tend to reduce the effective jet quenching parameter $\hat{q}$ with increasing virtuality in the \matter{} stage. In this effort, we do not include any such weakening effects. This is meant to produce an over-estimate of any energy loss effects from the \matter{} stage. 
There is also no LBT simulation, given the very short lifetime of the QGP in these systems. 

Due to the hard partons' interaction with the medium, the ability to track color exchanges throughout the shower is lost. 
Previously, in \jetscape{} simulations, one could use a colorless hadronization model, where the partonic color connects the final state partons with color assigned \emph{faux} remnants~\cite{Kumar:2019bvr}. 
However, for this study, we will contend with this limitation by employing a simple independent fragmentation model where the pion spectrum can be obtained from the following integral
\begin{align}
    \frac{dN^{\pi}}{dp_T}(p_T^{\pi})
    =& \sum_{f} \int \frac{dz}{z}~ D_{f\to \pi}(z) \frac{dN^f}{dp_T}\left(p_T^f = p_T^{\pi} / z\right),
\end{align}
where sum $f$ runs over all partonic flavors and $dN^f/dp_T$ is the partonic spectrum.
We use the KKP parametrization \cite{Kniehl:2000fe} for  the fragmentaiton function $D_{f\to \pi}(z)$.
Using a fragmentation function, has the added benefit that statistical error from the Pythia simulation that underlies the Colorless Hadronization module is now absent, yielding an even more accurate estimation of the shift in the hadronic $R_{\rm pPb}$ due to energy loss.  

In order to give the reader an even more physical picture of the simulation, we present a space-time picture of a typical event in an \xscapegim{} simulation. 
The initial position of the hard partons is determined from the hotspots of the \Glb{} model from which the hard scattering energy is subtracted.
In Fig.~\ref{fig:energy_profile}, we display the evolution of the transverse energy density profile of the bulk matter and the position of the hard partons from an example event.
One observes that while the hard partons started close to the center of the hydrodynamic evolution, after only $1$~fm/$c$ they have moved outside the freeze-out surface displayed in the red contour.

The partonic final state can be used to directly compute the jet spectrum using the anti-$k_T$ algorithm.
While partonic jets are known to not exactly reproduce hadronic jets observed in experiments, they nonetheless provide a close approximation to hadronic jets and can be used to obtain nuclear modification factors.
The results for nuclear modification factors using $N_{\rm scattering} \!=\! 6000$ and $N_{\rm shower} \!=\! 100$ are shown in Fig.~\ref{fig:PartonJetRpPb}.

We compare the partonic spectrum from a vacuum shower to a medium-modified shower using \matter{}. {In minimum bias \pPb{} collisions, experimental data provide evidence for a non-negligible amount of energy loss, with bounds placed at $\Delta E \lesssim 1\%$~\cite{ATLAS:2022iyq}, or $p_T \lesssim 0.4$~GeV out of jet cone~\cite{ALICE:2017svf}.}
Our results are consistent with these experiments, as the energy loss is negligible in both the jet and the pion spectra.
In Fig.~\ref{fig:PartonJetRpPb}, the nuclear modification factor $R_{\rm pPb}$ is consistent with unity within the statistical uncertainties.
In Fig.~\ref{fig:PionRpPb}, the hadronic $R_{\rm pPb}$ is larger than $0.9$. Thus, $|R_{\rm pPb} - 1|$ is within the typical uncertainty of simulations with millions of events (see Fig.~\ref{fig:RpPbhadron}), or experimental error bars.

In this section, we have demonstrated that even with the inclusion of final state energy loss, in the \matter{} module, there is no discernible  shift in the $R_{\rm pPb}$ for jets or charged hadrons from the case of no energy loss. The cause for this is straightforward, the produced bulk medium is simply too small and lives for too short a time (as shown in Fig.~\ref{fig:energy_profile}) to effect a noticeable shift in the $R_{\rm p Pb}$. Thus, it is possible for a QGP to be formed in small systems such as a $p-Pb$ collision without the appearance of any energy loss.

\section{Summary and Outlook}
\label{sec:summary}

The newly developed \xscape{} framework offers a comprehensive approach to study the interplay dynamics between hard and soft sectors in small systems, such as \pp{} and \pA{} collisions at high energies. Within this framework we have produced a simulator \xscapegim{} (\Glb{} + pythia + \imatter{} + \MUSIC{} + \matter{}) which allows for the production of small droplets of QGP, even in the presence of hard scattering, in \pp{} and \pA{} collisions. 

The \xscapegim{} simulator adopts a multi-stage approach and imposes soft and hard correlation, based on energy-momentum conservation, event by event. On the one hand, the soft non-perturbative particle production is modeled by the \Glb{} + hydrodynamics model. The \Glb{} model provides the collision geometry, including sub-nucleonic fluctuations, and parameterizes the initial bulk medium's energy-momentum distribution for the subsequent hydrodynamic evolution, after the initial collision. 
On the other hand, hard scatterings are sampled using Pythia and dressed with initial- and final-state radiation using the \imatter{} and \matter{} models. The energy and momentum consumed in the hard processes are subtracted from that available for bulk medium production. This multi-stage framework builds in non-trivial event-by-event correlations between soft and hard particle production in both spatial and momentum spaces. 

Using this hybrid approach, we presented a quantitative description of hadron and jet spectra in minimum-bias \pp{} and \pPb{} collisions across the full transverse momentum range at top LHC energies ($\sqrt{s_{\rm NN}} =5.02 {\rm TeV}$). We verified that the medium-induced energy loss is negligible in \pPb{} collisions, aligning with experimental findings. Future inclusion of shower-medium parton recombination is expected to reduce the tension at the intermediate $p_T$ range of the pion spectra. 

The soft-hard correlations from energy-momentum conservation unveiled an intriguing interplay between the high-$p_T$ trigger and the collision event activity, characterized by the transverse energy $E_T$ distribution in the forward (or backward) rapidity. The rise and fall of the system's transverse energy $E_T$ with the leading jet $p_T$ demonstrate how the hard scattering process produces an increasing $E_T$ over a wide range of rapidity (the pedestal), until energy momentum conservation limits and reduces the pedestal with increasing jet $p_T$. 
The \xscapegim{} results can qualitatively explain the correlations observed in the ALICE and ATLAS measurements.
The dynamically developed transition scale results from competition between the jet $p_T$ and partonic remnants. Future exploration of the collision energy dependence of correlations between event activity and hard process triggers at the top RHIC and LHC energies will shed light on understanding the dynamics in small systems.

The \xscape{} framework~\cite{x-scape-github} provides the community with a full-fledged event generator building toolkit for small systems to investigate the origin of collectivity and absence of medium-induced energy loss in high-multiplicity \pp{} and \pA{} collisions. Extending this framework to intermediate-size systems, such as $O$-$O$ collisions, and eventually large \AA{} collisions would further establish and verify this unified approach for relativistic nuclear collisions at high energy.

\section*{ACKNOWLEDGMENTS}

This work was supported in part by the National Science Foundation (NSF) within the framework of the JETSCAPE collaboration, under grant number OAC-2004571 (CSSI:X-SCAPE). It was also supported under  PHY-1516590 and PHY-1812431 (R.J.F., M.Ko., C.P. and A.S.); it was supported in part by the US Department of Energy, Office of Science, Office of Nuclear Physics under grant numbers \rm{DE-AC02-05CH11231} (X.-N.W. and W.Z.), \rm{DE-AC52-07NA27344} (A.A., R.A.S.), \rm{DE-SC0013460} (A.K., A.M., C.Sh., I.S., C.Si and R.D.), \rm{DE-SC0021969} (C.Sh. and W.Z.), \rm{DE-SC0024232} (C.Sh. and H.R.), \rm{DE-SC0012704} (B.S.), \rm{DE-FG02-92ER40713} (J.H.P. and M.Ke.), \rm{DE-FG02-05ER41367} (D.S. and S.A.B.), \rm{DE-SC0024660} (R.K.E), \rm{DE-SC0024347} (J.-F.P. and M.S.). The work was also supported in part by the National Science Foundation of China (NSFC) under grant numbers 11935007, 11861131009 and 11890714 (Y.H. and X.-N.W.), by the Natural Sciences and Engineering Research Council of Canada (C.G., S.J., and G.V.),  by the University of Regina President's Tri-Agency Grant Support Program (G.V.), by the Canada Research Chair program (G.V. and A.K.) reference number CRC-2022-00146, by the Office of the Vice President for Research (OVPR) at Wayne State University (Y.T.), and by the S\~{a}o Paulo Research Foundation (FAPESP) under projects 2016/24029-6, 2017/05685-2 and 2018/24720-6 (M.L.). C.Sh., J.-F.P. and R.K.E. acknowledge a DOE Office of Science Early Career Award. I.~S. was funded as a part of the European Research Council project ERC-2018-ADG-835105 YoctoLHC , and as a part of the Center of Excellence in Quark Matter of the Academy of Finland (project 346325).

Calculations for this work used the Wayne State Grid and Anvil at Purdue RCAC through allocation PHY230020 from the Advanced Cyberinfrastructure Coordination Ecosystem: Services \& Support (ACCESS) program~\cite{10.1145/3569951.3597559}, which is supported by National Science Foundation grants \#2138259, \#2138286, \#2138307, \#2137603, and \#2138296.
%Computations were carried out on the National Energy Research Scientific Computing Center (NERSC), a U.S.Department of Energy Office of Science User Facility operated under Contract No. DE-AC02-05CH11231. The bulk medium simulations were done using resources provided by the Open Science Grid (OSG) \cite{Pordes:2007zzb, Sfiligoi:2009cct}, which is supported by the National Science Foundation award \#2030508.
%Data storage was provided in part by the OSIRIS project supported by the National Science Foundation under grant number OAC-1541335.

\appendix
\section{Model parameters}
\label{App:ModelParams}

The \xscapegim{} simulator contains several model parameters to describe the initial state and the QGP properties in the bulk sector, as well as high-energy parton shower and their fragmentation. 

In this work, we vary 15 parameters summarised in Table~\ref{tab:ModelParams}. Their values were used in the simulation results presented in Sec.~\ref{sec:results}.

\begin{table}[h!]
    \centering
    \caption{Model parameters in \xscapegim{} simulator for small systems.}\label{tab:ModelParams}
    \begin{tabular}{c|c|c|c|c}
        \hline \hline
        $\hat{p}_{T, \mathrm{min}}$ & $p_{T,\mathrm{min}}$ & $\alpha_s$ & $C_v$ & {\texttt{probStoUD}} \\
        \hline
        8~GeV & 8~GeV & 0.15 & 0.25 & 0.217 \\
        \hline \hline
        
        $y_{\rm loss, 2}$ & $y_{\rm loss, 6}$ & $y_{\rm loss, 10}$ & $\rm \sigma_{y_{\rm loss}}$ & $\rm \sigma^{string}_{x}$ \\
        \hline
        1.6 & 2.15 & 2.15 & 0.5 & 0.5 fm \\
        \hline \hline
        
        $\rm \sigma^{string}_{\eta}$ & $\alpha_\mathrm{preflow}$ & $\eta/s$ & $\zeta_\mathrm{max}$ & $e_\mathrm{sw}$ \\
        \hline
        0.5 & 0.15 & 0.2 & 0.08 & 0.45 GeV/fm$^3$ \\
        \hline \hline
    \end{tabular}
\end{table}

In the hard sector, the parameter $\hat{p}_{T, \mathrm{min}}$ is the minimum invariant $p_T$ generating the initial hard collisions, and $p_{T, \mathrm{min}}$ is the miminal cut off for MPI. The strong coupling constant $\alpha_s$ is specified at the scale of $Z$ boson's mass and runs with parton's energy. The parameter $C_v$ determines the highest possible virtuality of initial partons in the \imatter{} and \matter{} parton shower.
In the Lund string fragmentation sector, the {\texttt{probStoUD}} describes the $s$ quark production probability relative to ordinary $u$ or $d$ one. 
We choose the non-perturbative virtuality scale $t_0 = 4$~GeV$^2$ for fragmentation in this work.

In the bulk sector, the rapidity loss parameters $y_{\mathrm{loss}, n}$ control the averaged  amount of energy loss for soft string production in the \Glb{} model using the following piece-wise parameterization~\cite{Shen:2023awv},
\begin{align}
    & \langle y_{\rm loss}\rangle(y_\mathrm{init}) = \nonumber \\
    & \quad \begin{cases}
        y_{\rm loss,2}\frac{y_{\rm init}}{2},\; 0<y_{\rm init}\leq 2\\
        y_{\rm loss,2} + (y_{\rm loss,6} - y_{\rm loss,2})\frac{y_{\rm init}-2}{4},\; 2<y_{\rm init}<6\\
        y_{\rm loss,6} + (y_{\rm loss,10} - y_{\rm loss,6})\frac{y_{\rm init}-6}{4},\; y_{\rm init}\geq 6
    \end{cases}
    \label{eq:ylossParam}
\end{align}
Here, $y_\mathrm{init}$ is the magnitude of the incoming parton rapidity in the collision pair rest frame.
The rapidity loss fluctuation is introduced by the variance parameter $\sigma_{y_\mathrm{loss}}$ with the logit-normal distribution. We sample a random number $X$ from a normal distribution $\mathcal{N}(0, \sigma_{y_\mathrm{loss}})$ and compute $Y \equiv 1/(1 + e^{-X})$. Then the rapidity loss $y_\mathrm{loss}(y_\mathrm{init})$ is computed as,
\begin{align}
   y_\mathrm{loss}(y_\mathrm{init}) &= (4 \langle y_{\rm loss}\rangle - y_\mathrm{init})Y + 2(y_\mathrm{init} - 2 \langle y_{\rm loss}\rangle) Y^2.
   \label{eq:ylossSample}
\end{align}
The sampled rapidity loss $y_\mathrm{loss}$ from Eq.~\eqref{eq:ylossSample} is bounded between 0 and incoming rapidity $y_\mathrm{init}$ and has a mean $\langle y_\mathrm{loss} \rangle$~\cite{Shen:2022oyg}. 

The parameter $\sigma_x^\mathrm{string}$ is the Gaussian width for the hotspot's transverse spatial profile for the energy-momentum source terms. And the $\sigma_\eta^\mathrm{string}$ controls the Gaussian fall off at the string ends along the space-time rapidity direction.

When soft strings are deposited to hydrodynamic fields, they develop pre-equilibrium transverse velocity fields parameterized by the blast-wave profile~\cite{Zhao:2022ugy}. For a soft string at transverse position $(x_\mathrm{string}, y_\mathrm{string})$, the transverse flow rapidity is parameterized as,
\begin{equation}
    \eta_\perp (\textbf{x}_\perp) = \alpha_\mathrm{preflow} \vert \tilde{\textbf{x}}_\perp \vert,
\end{equation}
where $\tilde{\textbf{x}}_\perp = (x - x_\mathrm{string}, y - y_\mathrm{string})$ denotes the 2D vector points from the string center. Then, the 2D transverse flow vector is $\textbf{u}_\perp (\textbf{x}_\perp) = \sinh[\eta_\perp(\textbf{x}_\perp)] \hat{\textbf{e}}_{\tilde{\textbf{x}}_\perp}$ with $\hat{\textbf{e}}_{\tilde{\textbf{x}}_\perp} = \tilde{\textbf{x}}_\perp/\vert \tilde{\textbf{x}}_\perp \vert$ being the unit vector points along the $\tilde{\textbf{x}}_\perp$ direction. 

We consider shear and bulk viscous effects in hydrodynamic evolution. The QGP-specific shear viscosity is chosen to be an effective constant, $\eta T/(e+P) = 0.2$, and the temperature-dependent QGP-specific bulk viscosity is parameterized as~\cite{Zhao:2022ugy, Shen:2023awv},
\begin{equation}
    \frac{\zeta T}{(e+P)}(T)=\zeta_{\rm max}{\rm exp}\left[-\left(\frac{T-T_{\rm peak}}{T_{{\rm width},\lessgtr}}\right)^2\right],
    \label{eq:bulk}
\end{equation}
where $T_{\rm width,<} = 0.015$\,GeV for $T<T_{\rm peak}$, and $T_{\rm width, >} = 0.1$\,GeV for $T> T_{\rm peak}$ with $T_{\rm peak}=0.16$ GeV~\cite{Zhao:2022ugy}.

The parameter $e_\mathrm{sw}$ is the switching energy density for converting fluid cells to hadrons via the Cooper-Frye particlization in the bulk sector.
\begin{figure*}
    \centering
    \includegraphics[width=0.9\textwidth]{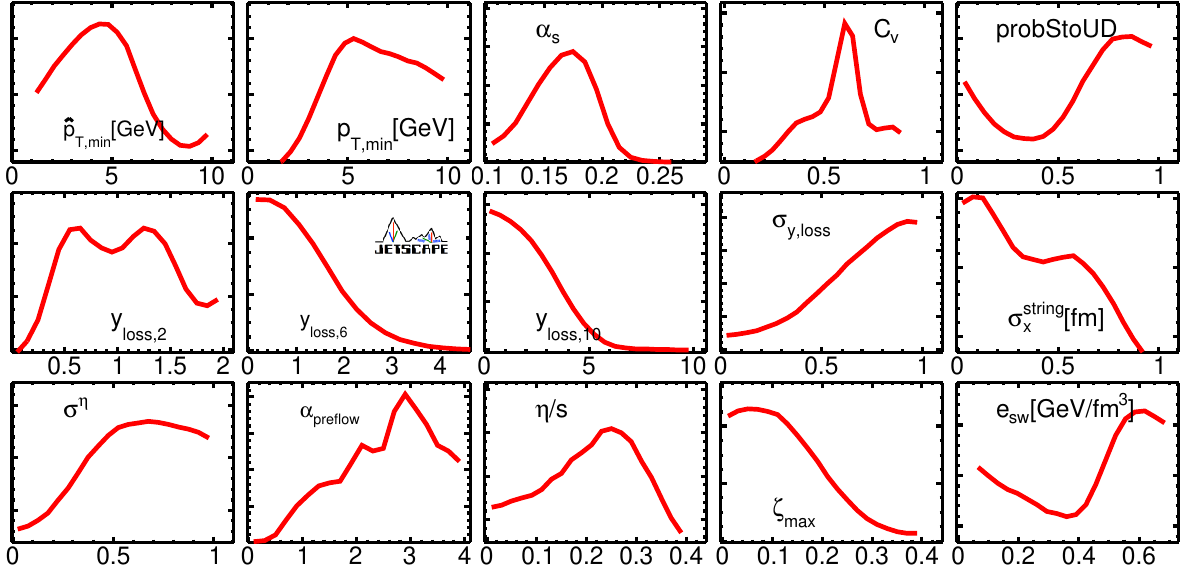}
    \caption{The normalized marginal distributions for
individual parameters from the posterior distributions of the Bayesian inference analysis. The ranges in the $x$-axis indicate the uniformed prior ranges for each model parameter.}
\label{fig:posterior}
\end{figure*}

We carried out the first iteration of the Bayesian calibration for the 15 model parameters with a limited training design in Sec.~\ref{sec:Bayesian}. The marginal distributions for each parameter are shown in Fig.~\ref{fig:posterior}. With only the identified particle $p_T$-spectra in minimum bias \pp{} collisions, we get some constraints on the parameters in the hard sector. The model parameters in the bulk sectors are weakly constrained. Our current study sets up the Bayesian workflow for a more dedicated study in the future.

\bibliography{inspire, non-inspire}
\end{document}